\documentclass[12pt,a4paper,twoside]{article}

\usepackage[utf8]{inputenc}
\usepackage{dirtytalk}
\usepackage{tikz}
\usepackage{float}
\usepackage{tkz-tab}
\usepackage{amsmath}
\usepackage{amssymb} 
\usepackage{tikz}
\usepackage{float}
\usepackage{bm}
\usepackage{amsthm}
\usepackage{fancyhdr}
\usepackage{enumitem}
\usepackage{mathrsfs} 
\usepackage{graphicx}
\usepackage{xcolor}
\usepackage{color}
\usepackage{dsfont}
\usepackage{yfonts}
\usepackage{fancybox}
\usepackage{listings}
\usepackage{lineno}
\newcommand{\f}{\dfrac}

\newcommand{\hp}{\hat p}
\newcommand{\hc}{\hat c}
\newcommand{\ome}{\omega}
\newcommand{\uq}{\frac{1}{4}}

\newcommand{\dpd}{2 \pi \delta}
\newcommand{\eps}{\epsilon}
\newcommand{\beq}{\begin{equation}}
\newcommand{\eeq}{\end{equation}}
\newcommand{\beqn}{\begin{equation*}}
\newcommand{\eeqn}{\end{equation*}}
\newcommand{\bal}{\begin{align}}
\newcommand{\eal}{\end{align}}
\newcommand{\baln}{\begin{align*}}
\newcommand{\ealn}{\end{align*}}

\begin{document}

\lstset{ numbers=left, tabsize=3, frame=single, numberstyle=\ttfamily, basicstyle=\footnotesize} 
\thispagestyle{empty}

\begin{center}
\shadowbox{
\begin{minipage}{1\textwidth}
\begin{center}
{\Huge Pressure beneath a periodic travelling water-wave in constant-vorticity flow over a flat bed}\\
\end{center}
\end{minipage}
}\\
\begin{center}
	A. Constantin$^1$ \let\thefootnote\relax\footnotetext{$^1$Faculty of Mathematics
University of Vienna
Oskar-Morgenstern-Platz 1
	A-1090 Vienna, Austria}, N. Gindrier$^2$ \let\thefootnote\relax\footnotetext{$^2$Johann Radon Institute for Computational
	and Applied Mathematics (RICAM)
	Altenbergerstraße 69
	A-4040 Linz, Austria}, O.Scherzer$^{1,2,3}$\let\thefootnote\relax\footnotetext{$^3$ Christian Doppler Laboratory
for Mathematical Modeling and Simulation
of Next Generations of Ultrasound Devices (MaMSi)
Oskar-Morgenstern-Platz 1
A-1090 Vienna, Austria} \\
\end{center}
\end{center}

\begin{abstract}
We investigate within the framework of linear theory the behaviour of the total (hydrodynamic) pressure and of the dynamic pressure in a regular wave train which propagates at the surface of water with a flat bed 
in a flow with constant vorticity. We show that nonzero vorticity, the hallmark of a non-uniform underlying current, may strongly alter the behaviour with respect to the case of irrotational flows, for which the 
maximum and minimum of the dynamic pressure always occur at the wave crest and at the wave trough, respectively -- the extrema of the dynamic pressure may occur along the flat bed or along the critical level, depending on the 
vorticity strength. While vorticity does not modify the increase of the hydrodynamic pressure with depth, it can 
significantly alter the location of the extrema of the hydrodynamic pressure at a fixed depth level.
\end{abstract}

\section{Introduction}

The estimation of underwater extreme pressures has great importance in assessing the associated impact
on structures. Subsurface and bottom pressure sensors are commonly used to measure surface waves. 
Bottom pressure sensors often estimate wave heights by assuming a hydrostatic pressure distribution but 
non-hydrostatic corrections to the pressure are sometimes significant (see \cite{ba, cb, cl, cla, cc, ch, clh}). 

Most waves that propagating on the surface of oceans, seas, lakes or canals are caused by wind. Some of these 
waves are very localized (for example, ripples)  but gravity waves (for which surface tension is negligible, the dominant restoring force being gravity) 
often have enough energy to travel well beyond the place of their origin. When they move away from their generation zone, waves become more regular and are referred to as swell. 
Such two-dimensional periodic wave trains can travel with practically constant speed and no change of direction or shape over distances measured in hundreds of km (see \cite{k}). In the absence of 
underlying currents, their flow is irrotational but sometimes they interact with non-uniform currents. Detailed studies of the pressure beneath irrotational travelling waves are available 
(see the discussion in \cite{c-jpa, c-pf, cs, ovdh, u, z,es,ko}) but the case of wave-current interactions is less well-understood (see the discussion in \cite{ht}). We will investigate two-dimensional inviscid and periodic travelling waves propagating over a flat bed in 
a flow with constant vorticity. The inviscid setting is realistic since the velocity
profile in the water, whether due to laminar viscosity or turbulent mixing, is usually
established over timescales/lengthscales which are very long compared with a wave period/wavelength. As for the choice of constant vorticity, note that this is considered adequate for wind-generated waves 
(see \cite{e}). Furthermore, when the waves are long compared with the water depth, it is the existence of a non-zero mean vorticity that is important rather than its
specific distribution (see \cite{dp}). Another assumption that is very convenient for our purposes is that the water bed is horizontally flat. This is the case for water waves propagating in a canal, as well as 
for surface waves in sea regions with a flat bed. For example, abyssal plains (vast sediment-covered regions of the sea floor that are the flattest areas on
Earth,  presenting variations in depth below 1m per km of horizontal distance, resulting from the blanketing of a preexisting
irregular ocean floor topography by accumulated land-derived sediment) are found
in all major sea and ocean basins, at depths between 2 and
6 km, covering overall almost a third of the Earth’s surface (about as much as all the exposed land combined). Another important example is provided by a shallower sea region with a practically flat bed -- 
the continental shelf, sloping gently toward the deep ocean basins at approximately 0.5 to 1 degree angles. The continental shelf reaches a width in excess of 1000 km in Siberia (see \cite{m}) but is 
practically absent on the French Riviera.

With the ultimate goal to find the location of the extrema of the hydrodynamic and of the dynamic pressure, the paper is organized as follows. In Section 2 we present the governing equations and the nondimensionalisation process necessary to obtain the associated linearization. We also discuss 
the dispersion relation which selects the physically relevant solutions, representing waves of small amplitude. The possible appearance of critical layers is also addressed. Section 3 is devoted 
to the study of the location of the extrema of the dynamic pressure, according to the values of the vorticity. The total pressure is investigated in Section 4. We conclude with a summary of our findings in Section 5.

\section{Preliminaries}

We now briefly present the governing equations and discuss their linearization which allows us to perform an in-depth study of waves of small amplitude.

\subsection{Governing equations}

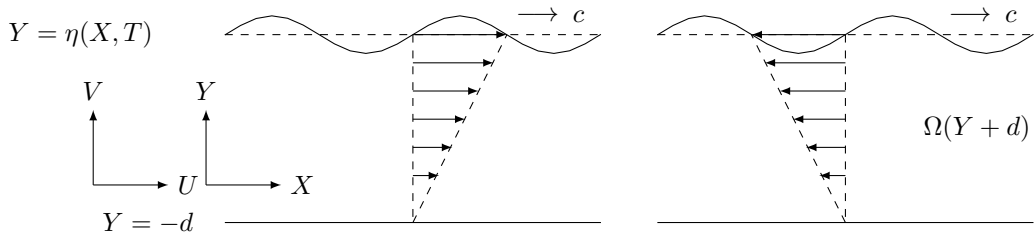
\begin{figure}[htbp]
\centering
\begin{tikzpicture}[scale=2.5]
\draw(-1,-1)--(1,-1);
\draw[dashed](0,0)--(0,-1);
\draw [domain=-1:1] plot (\x,{sin(6.3*\x r)*0.1});
\draw[dashed](-1,0)--(1,0);
\draw(-1.7,-1) node[right]{$Y=-d$};
\draw(-2.2,0) node[right]{$Y=\eta(X,T)$};
	\draw[->,>=latex](0,0)--+(0.5,0);
	\draw[->,>=latex](0,-0.15)--+(0.43,0);
	\draw[->,>=latex](0,-0.3)--+(0.35,0);
	\draw[->,>=latex](0,-0.45)--+(0.28,0);
	\draw[->,>=latex](0,-0.6)--+(0.21,0);
	\draw[->,>=latex](0,-0.75)--+(0.14,0);
	\draw[dashed](0,-1)--(0.5,0);
\draw[->,>=latex](-1.1,-0.8)--+(0,0.4) node[above]{$Y$};
\draw[->,>=latex](-1.1,-0.8)--+(0.4,0) node[right]{$X$};
\draw[->,>=latex](-1.7,-0.8)--+(0,0.4) node[above]{$V$};
\draw[->,>=latex](-1.7,-0.8)--+(0.4,0) node[right]{$U$};
\draw(0.8,0.1) node[right]{$c$};
\draw(0.5,0.1) node[right]{$\longrightarrow$};
	\begin{scope}[xshift=2.3cm]
\draw(-1,-1)--(1,-1);
\draw[dashed](0,0)--(0,-1);
\draw [domain=-1:1] plot (\x,{sin(6.3*\x r)*0.1});
\draw[dashed](-1,0)--(1,0);
\draw(0.8,0.1) node[right]{$c$};
\draw(0.5,0.1) node[right]{$\longrightarrow$};
	\draw[->,>=latex](0,0)--+(-0.5,0);
	\draw[->,>=latex](0,-0.15)--+(-0.43,0);
	\draw[->,>=latex](0,-0.3)--+(-0.35,0);
	\draw[->,>=latex](0,-0.45)--+(-0.28,0);
	\draw[->,>=latex](0,-0.6)--+(-0.21,0);
	\draw[->,>=latex](0,-0.75)--+(-0.14,0);
	\draw[dashed](0,-1)--(-0.5,0);
	\draw(0.7,-0.5) node{$\Omega(Y+d)$};
\draw(0,0) node[right]{};
\draw(0,-1) node[right]{};
\end{scope}
\end{tikzpicture}
	\caption{Depiction of a wave propagating to the right with velocity $c>0$ and an underlying current with velocity $\Omega(Y+d)$ vanishing at the flat bed $Y=-d$.  \textit{Left}: $\Omega>0$. \textit{Right}: $\Omega<0$.}
\label{fig:model}
\end{figure}

We consider a two-dimensional wave travelling with velocity $c>0$, in water with a flat bed $Y=-d$ where the flow vanishes. We assume the presence of an underlying current $(\Omega(Y+d),0)$ of constant vorticity $\Omega$, called  
\textit{favourable} if $\Omega >0$ and \textit{adverse}  if $\Omega<0$. Fig. \ref{fig:model} illustrates the configurations with the two possible directions of the underlying current. We regard the flow field beneath the surface wave and over the flat 
bed $Y=-d$ as a wave-current interaction with velocity field
\begin{linenomath*}
$$(U(X,Y,T)+\Omega(Y+d),V(X,Y,T))\,.$$
\end{linenomath*}

The flow is modelled by the free-boundary homogeneous Euler equations with gravity as the restoring force. Consequently we have the following governing equations in the fluid domain delimited below by the flat bed $Y=-d$ and above by the free 
surface $Y=\eta(X,T)$, with $X$ oriented in the direction of wave propagation and with $Y$ vertically upward, the mean surface level being $Y=0$:
\begin{itemize}
\item Mass conservation: the velocity field $(U,V)$ satisfies
\beq 
U_X+V_Y=0,\hspace{3cm}-d \leq Y \leq \eta(X,T) \,;
\label{mc0}
\eeq
\item Euler equations:
\begin{linenomath*}
\begin{equation}\label{ee0}
\left\{\begin{array}{l}
U_T + [\Omega(Y+d)+U]U_X +V[\Omega +U_Y]= - \displaystyle\frac{1}{\rho}\,P_X\,,\\[0.3cm]
V_T +[\Omega(Y+d)+U]V_X +VV_Y =-\displaystyle\frac{1}{\rho}\,P_Y\,-g\,,
\end{array}\right. 
\qquad -d \le Y \le \eta(X,T)
\end{equation}
\end{linenomath*}
where $P$ is the pressure, $\rho$ is the (constant) density, $g$ is the (constant) gravitational acceleration and
\beq 
0=U_Y-V_X,\hspace{3cm}-d \leq Y \leq \eta(X,T) 
\label{vort0}
\eeq
because constant vorticity $\Omega$ the wave perturbation is irrotational.
\end{itemize}

The associated boundary conditions are the impermeability of the flat, rigid bed
\begin{linenomath*}
\begin{equation}\label{bcb0}
V=0\quad\hbox{on}\quad Y=-d\,,
\end{equation}
\end{linenomath*}
and the kinematic condition 
\begin{linenomath*}
\begin{equation}\label{bcs0}
V=\eta_T + [\Omega(Y+d)+U]\eta_X\quad\hbox{on}\quad Y=\eta(X,T)\,,
\end{equation}
\end{linenomath*}
on the free air-water interface, and the dynamic boundary condition
\begin{linenomath*}
\begin{equation}\label{bcp0}
P=P_{atm}\quad\hbox{on}\quad Y=\eta(X,T)\,,
\end{equation}
\end{linenomath*}
with $P_{atm}$ the (constant) atmospheric pressure, which decouples the motion of the water from that of the air above it.

We introduce the dynamic pressure ${\frak P}$ by means of 
\begin{linenomath*}
\begin{equation}\label{pn}
P(X,Y,T)=P_{atm} - \rho g Y + \rho g d\, {\frak P}(X,Y,T),\qquad -d \le Y \le \eta(X,T)\,.
\end{equation}
\end{linenomath*}

For recent analytical and numerical studies of the governing equations we refer to \cite{cvs,cvs2,i,ka,ks,l,lww,rmn,w,w2}.

\subsubsection{Nondimensionalisation}
\label{sec:nonde}
\paragraph{Notations}
To linearize it is necessary to write the governing equations in nondimensional form. With this purpose, we denote the wavelength by $L$, the water mean depth by $d$, the typical wave amplitude by $a$ and introduce 
the  shallowness parameter
\begin{linenomath*}
\begin{equation}\label{delta}
\delta=\frac{d}{L}
\end{equation}
\end{linenomath*}
and the amplitude parameter
\begin{linenomath*}
\begin{equation}\label{eps}
\varepsilon=\frac{a}{d}\,.
\end{equation}
\end{linenomath*}

There are two fundamental length scales, to account for the horizontal and vertical directions. In addition, there are speed scales in each of these 
directions, and an amplitude scale (if waves are present) with an associated time scale. The relative sizes of these scales lead to 
corresponding approximations of the original system and to the construction of asymptotic expansions. Suitable nondimensional variables are defined in terms of the wave speed scale $c_0=\sqrt{gd}$ characteristic for 
irrotational shallow water waves (see \cite{cb}) by
\begin{linenomath*}
\begin{equation}\label{nd}
X=L\, x\,,\quad Y=d\,y\,,\quad T=\displaystyle\frac{L}{c_0}\,t\,,\quad U=c_0\,u\,,\qquad V=\displaystyle\frac{dc_0}{L}\,v\,,
\end{equation}
\end{linenomath*}
so that
$$U_X=\f{c_0}{L}u_x\,,\qquad U_Y=\f{c_0}{d}u_y \,,\qquad U_T=\f{u_t}{L} \,,\qquad V_Y=\f{c_0}{L}v_y\,,\qquad V_T=d v_t \f{c_0^2}{L^2}\,.$$
The rationale for the above non-dimensionalisation of the vertical velocity component is the consistency with 
the equation of mass conservation: setting $V=\alpha v$, (\ref{mc0}) becomes 
\begin{linenomath*}
$$\frac{c_0}{L}\,u_x + \frac{\alpha}{d}\,v_y=0\,,$$
\end{linenomath*}
and imposing the preservation of the divergence form of the equation for mass conservation to ensure the existence of a stream function 
leads to $\alpha=c_0d/L$. With this choice, the constraint (\ref{vort0}) becomes 
\begin{linenomath*}
\begin{equation}\label{ifnd}
u_y=\delta^2\,v_x\,,
\end{equation}
\end{linenomath*}

Now, setting
\begin{linenomath*}
\begin{equation}\label{snd}
\eta(X,T)=a\,h(x,t)\,,
\end{equation}
\end{linenomath*}
we transform the governing equations (\ref{mc0})-(\ref{ee0})-(\ref{vort0})-(\ref{bcb0})-(\ref{bcs0})-(\ref{bcp0}) into
\begin{linenomath*}
\begin{equation}\label{gend}
\begin{array}{l}
u_x+v_y=0\quad\hbox{for}\quad -1 < y < \varepsilon\,h\,,\\[0.1cm]
\left\{\begin{array}{l}
u_t+[u+\omega(1+y)]u_x+v[\omega +u_y]=-\,p_x \,,\\[0.25cm]
\delta^2\Big(v_t+[u+\omega(1+y)]v_x+vv_y\Big)=-\,p_y\,,
\end{array}\right. 
\quad\hbox{for}\quad -1 < y < \varepsilon\,h\,,\\[0.1cm]
v=0\quad\hbox{on}\quad y=-1\,,\\[0.1cm]
v=\varepsilon\Big(h_t+[u+\omega(1+\varepsilon h)]h_x\Big) \quad\hbox{on}\quad y=\varepsilon \,h\,,\\[0.1cm]
p=\varepsilon\,h \quad\hbox{on}\quad y=\varepsilon \,h\,,\\[0.1cm]
u_y=\delta^2 \,v_x\quad\hbox{for}\quad -1 < y < \varepsilon\,h\,,
\end{array}
\end{equation}
\end{linenomath*}
where 
\beq
\omega=\Omega\,\sqrt{ \f{d}{g}}
\label{ssnd}
\eeq
and
\begin{linenomath*}
\begin{equation}\label{pnp}
p(x,y,t)={\frak P}(X,Y,T)\,.
\end{equation}
\end{linenomath*}

\paragraph{Reference values}
\label{sec:approx}

It is of interest to provide reference values for the physically relevant parameters:
\begin{itemize}
	\item $\rho=1025$ kg m$^{-3}$ is the average density of sea water. The density varies with temperature, salinity, and pressure, but always exceeds the density 1000 kg m$^{-3}$ of fresh water at 4$^\circ$ C and 
	the highest value recorded by Britannica is 1071.04 kg m$^{-3}$ at 10 km depth.
	\item $g=9.81$ m s$^{-2}$ is the gravitational acceleration at sea level, the change throughout the ocean depth is negligible
	\item $P_{atm}=10^5$ $\text{kg}\,\text{m}^{-1}\text{s}^{-2}$ (Pa) is the typical atmospheric pressure at sea level, with variations confined to the range 93\%-106\% according to the NOAA database
	\item $1\,\text{m} \leq d \leq 6000$ m, with the lower range adequate for canals (for example, Londons Regent's Canal, is about 1 m deep, 5 m wide and 14 km long, while the Canal du Midi in France is about 1.5 m deep 
	and 240 km long, of average width 20 m) and the upper range for abyssal plains (for example, the Argentine Abyssal Plain Atlantic, located at 6 km depth, with a diameter of about 1000 km). 
	The shallowest continental shelf is found on the bed of the 
	Eastern Siberia Sea, with an average depth of 20 m, extending more than 1000 km offshore, along a continental coastline more than 1000 km long. 
	\item $3$ m $< L< 400$ m, with ocean waves rarely having wavelengths greater than 200 m (see \cite{k}). Wind waves in canals can be just 3 m long due to their confined environment.
	\item $|\Omega| \le 1$ s$^{-1}$, the highest value occurring in the special circumstances of the strong tidal current along the 30 m deep, 3km long and 150m wide Saltstraumen channel, north of the Arctic Circle on the coast of Norway 
	(see \cite{eli}). On the other hand, the vorticity of ocean currents does not exceed 0.05 s$^{-1}$. For example, the vorticity of the tidal currents in the about 35 deep Pentland Firth between the 
	Scottish mainland and the Orkney Islands reaches 0.03 s$^{-1}$ (see \cite{ck}), while the jets of the Antarctic Circumpolar Current attain a surface speed of 1 m s$^{-1}$ 
	and extend down to the 4 km deep floor of the Southern Ocean, with a vorticity $\Omega \approx 2.5 \times 10^{-4}$ s$^{-1}$ (see \cite{ac}). 
	In this context, note the empirical formula $|\Omega|=0.02 \times \frac{W_{10}}{d}$, widely used in ocean engineering for wind-drift currents reaching down to the bed, $W_{10}$ being the wind speed at 10 m 
	above sea level (see \cite{e}). According to this formula, for $d \ge 20$ m, winds faster than 50 m s$^{-1}$ are required for the vorticity to exceed 0.05 s$^{-1}$ and persistent wind 
	speeds in excess 25 m s$^{-1}$ are rare even in the Southern Ocean, 
	where the strongest average winds on Earth occur. As a further illustration of the empirical formula, consider the Transpolar Drift current in the Arctic Ocean, flowing away from the continental shelf of the Eastern Siberia Sea 
	across the North Pole towards the Fram Strait. In summer there is no ice cover of the Eastern Siberia Sea and 2-3 m high waves with wavelengths of the order of 60–90 m propagate at the water surface (see \cite{m}). 	
	Data from the Interim European Reanalysis shows that the 10 m winds in the summer in this region may exceed 10 m s$^{-1}$ for several hours, and the formula yields $\Omega \approx 10^{-2}$ s$^{-1}$. 
	\item $0.5$ m $<a< 30$ m, the average ocean wave height being less than 2 m but with waves in the Southern Ocean often more than 6 m high, while in 
	1995 waves with amplitudes in excess of 25 m were detected near oil platforms in the North Sea. 
\end{itemize}
Note that the values of $\epsilon$ and $\delta$ are computed from (\ref{eps}) and \eqref{delta}, respectively, with the linearization process requiring $\epsilon \ll 1$, while $\delta \ll1$ defines the shallow-water regime.

\subsubsection{Linearization and travelling waves}
To derive the linear equations governing the propagation of waves of small amplitude, we proceed in three steps:
\begin{enumerate}
	\item we first perform the scaling $u=\eps \hat u$, $v=\eps \hat v$, $p=\eps \hp$ (see \ref{scale})
	\item as the outcome of the limiting process $\eps \rightarrow 0$, we obtain the linearization
	\item we make the travelling-wave {\it Ansatz} by requiring an $(x,t)$-dependence of the form 
	$$x-\hat{c}t = \f{X-\hat{c}\sqrt{gd}T}{L}$$
	with $c=\hat c \sqrt{gd}$.
		\end{enumerate}
Note that $\eps \rightarrow 0$ means that the amplitude of the waves is negligible compared to depth. Regarding the solutions  as 
wave perturbations of the underlying current, it is appropriate to envisage an asymptotic expansion in 
powers of $\varepsilon$. The fact that the absence of waves corresponds to setting $\varepsilon=0$ and 
$u \equiv 0$, $v \equiv 0$, $p \equiv 0,$ in (\ref{gend}), suggests the scaling
\begin{linenomath*}
\begin{equation}\label{scale}
u(x,y,t)=\varepsilon\,\hat{u}(x,y,t),\  v(x,y,t)=\varepsilon\,\hat{v}(x,y,t),\quad  p(x,y,t)=\varepsilon\,\hat{p}(x,y,t)\,.
\end{equation}
\end{linenomath*}
A scaling of $h$ is not necessary, since by (\ref{snd}) this function already arises as the scaled deviation 
of the wave profile from the flat state of the surface of still water. 
 From (\ref{gend}), (\ref{ssnd}) and (\ref{scale}), the full set of dimensionless, scaled governing equations and boundary conditions become
\begin{linenomath*}
\begin{equation}\label{gens}
\begin{array}{l}
\hat{u}_x+\hat{v}_y=0\quad\hbox{for}\quad -1 < y < \varepsilon\,h\,,\\[0.1cm]
\left\{\begin{array}{l}
\hat{u}_t+[\omega(1+y)+ \varepsilon \hat{u}]\hat{u}_x+\hat{v}[\omega +\varepsilon\hat{u}_y]=-\,\hat{p}_x \,,\\[0.2cm]
\delta^2\Big(\hat{v}_t+[\omega(1+y)+ \varepsilon \hat{u}]\hat{v}_x+\varepsilon\hat{v}\hat{v}_y\Big)=-\,\hat{p}_y\,,
\end{array}\right. 
\quad\hbox{for}\quad -1 < y < \varepsilon\,h\,,\\[0.1cm]
\hat{v}=0\quad\hbox{on}\quad y=-1\,,\\[0.1cm]
\hat{v}=h_t+[\omega(1+\varepsilon \,h)+ \varepsilon \hat{u}]h_x \quad\hbox{on}\quad y=\varepsilon \,h\,,\\[0.1cm]
\hat{p}=h \quad\hbox{on}\quad y=\varepsilon \,h\,,\\[0.1cm]
\hat{u}_y=\delta^2 \hat{v}_x\quad\hbox{for}\quad -1 < y < \varepsilon\,h\,.
\end{array}
\end{equation}
\end{linenomath*}

The three quantities (typical amplitude $a$, typical wavelength $L$ and mean depth $d$) involved, 
by means of (\ref{delta}) and (\ref{eps}), in the re-formulation (\ref{gens}) of the governing equations are independent. 
Consequently, regarding the underlying current of constant vorticity as a permanent feature of the flow, independent 
of the presence of waves, the free-boundary problem (\ref{gens}) contains two independent parameters, $\varepsilon$ and $\delta$, 
measuring a typical amplitude and a typical wavelength relative to the average depth of the water. Due to \eqref{pn} and 
\eqref{pnp}, the pressure $P(X,Y,T)$ beneath the waves is given in physical units by
\begin{linenomath*}
\begin{equation}\label{presu}
P_{atm} - \rho gd\, y + \varepsilon \rho gd \,\hat{p}(x,y,t)\,,
\end{equation}
\end{linenomath*}
expression in which $\hat{p}(x,y,t)$ is the nondimensional dynamic pressure, while $P_{atm} - \rho gd\, y$ is the hydrostatic pressure. Note that in the absence of waves ($\varepsilon=0$) the 
pressure is hydrostatic, irrespective of the strength $\Omega$ of the underlying linearly sheared current.

The linearization corresponds to $\varepsilon \to 0$ (with $\delta$ fixed) and 
consists of the set of equations
\begin{linenomath*}
\begin{equation}\label{lin}
\left\{\begin{array}{l}
\hat{u}_x+\hat{v}_y=0\quad\hbox{for}\quad -1 < y < 0\,,\\[0.1cm]
\hat{u}_t + [\omega(1+y)]\,\hat{u}_x + \omega \hat{v}=-\,\hat{p}_x \quad\hbox{for}\quad -1 < y < 0\,,\\[0.1cm]
\delta^2\,\Big(\hat{v}_t + [\omega(1+y)]\hat{v}_x\Big)=-\,\hat{p}_y\quad\hbox{for}\quad -1 < y < 0\,,\\[0.1cm]
\hat{v}=0\quad\hbox{on}\quad y=-1\,,\\[0.1cm]
\hat{v}=h_t +\omega\, h_x\quad\hbox{on}\quad y=0\,,\\[0.1cm]
\hat{p}=h \quad\hbox{on}\quad y=0\,,\\[0.1cm]
\hat{u}_y=\delta^2 \,\hat{v}_x\quad\hbox{for}\quad -1 < y < 0\,.
\end{array}\right.
\end{equation}
\end{linenomath*}
This deceptively simple system presents a peculiar feature: the evaluation on the free surface from (\ref{gens}) is 
replaced by an evaluation on the known surface $y=0$, yet the free surface profile --- which is the primary unknown --- 
still appears in the problem by means of the boundary conditions on $y=0$.

To analyze the linearised system (\ref{lin}) we restrict our attention to periodic travelling waves of unit 
spatial period and speed $\hat{c}>0$ in the non-dimensional set-up, corresponding to the wavelength $L$ and the wave 
speed $c=\hat{c}\sqrt{gd}$ in physical variables, since
\begin{linenomath*}
$$x-\hat{c}\,t=\frac{X-\hat{c}\sqrt{gd}\,T}{L}$$
\end{linenomath*}
by \eqref{nd} and \eqref{ssnd}. In this setting $\hat u_t=-c\hat u_x$ and $h_t=-c h'(x)$. With an 
$(x,t)$-dependence of $\hat{u}$, $\hat{v}$, $\hat{p}$, and $h$ of the form $(x-\hat{c}\,t)$, corresponding to describing the flow in a frame of reference moving at the wave speed, the system 
(\ref{lin}) becomes 
\begin{linenomath*}
\begin{equation}\label{lint}
\left\{\begin{array}{l}
\hat{u}_x(x,y)+\hat{v}_y(x,y)=0\quad\hbox{for}\quad -1 < y < 0\,,\\[0.1cm]
[\hat{c}  -\omega(1+y)]\,\hat{u}_x(x,y)- \omega \hat{v}(x,y)=\hat{p}_x(x,y) \quad\hbox{for}\quad -1 < y < 0\,,\\[0.1cm]
[\hat{c} - \omega(1+y)]\,\delta^2\,\hat{v}_x(x,y)=\hat{p}_y(x,y)\quad\hbox{for}\quad -1 < y < 0\,,\\[0.1cm]
\hat{v}(x,-1)=0\quad\hbox{on}\quad y=-1\,,\\[0.1cm]
\hat{v}(x,0)=[\omega-\,\hat{c}]\,h'(x)\quad\hbox{on}\quad y=0\,,\\[0.1cm]
\hat{p}(x,0)=h(x) \quad\hbox{on}\quad y=0\,,\\[0.1cm]
\hat{u}_y(x,y)=\delta^2 \,\hat{v}_x(x,y)\quad\hbox{for}\quad -1 < y < 0\,.
\end{array}\right.
\end{equation}
\end{linenomath*}

\subsubsection{The dispersion relation}

Within the framework of linear theory we may decompose the surface profile $h(x)$ of a non-dimensional travelling wave with unit period and wave speed $c$ in the Fourier series
\begin{linenomath*} 
$$h(x)=\sum_{k \ge 1} h_k\,\sin(2\pi kx)\,,$$
\end{linenomath*} 
since the mean water level is $y=0$. The dispersion relation
\begin{linenomath*}
\begin{equation}\label{eqfc2}
\hat{c}-\omega =-\,\frac{\omega\tanh(2\pi k\delta)}{4\pi k \delta} \pm \sqrt{\frac{\omega^2\tanh^2(2\pi k \delta)}{(4\pi k \delta)^2} + 
\frac{\tanh(2\pi k \delta)}{2\pi k \delta}}\,,
\end{equation}
\end{linenomath*}
specifies the possible wave speeds: given the shallowness parameter $\delta>0$, for any integer $k \ge 1$ we have only two possible wave speeds $\hat{c}_k$, one smaller than 
$\omega$ and one larger than $\omega$, with no superposition of Fourier modes (see Appendix 1). Without loss of generality, we consider travelling wave solutions with principal unit period, having a surface wave of the form 
\begin{linenomath*}
\begin{equation}\label{lw}
h(x)= A\,\sin(2\pi x)\,.
\end{equation}
\end{linenomath*}
The corresponding velocity field is given by
\begin{linenomath*}
\begin{align*}
&\hat{v}(x,y)= \gamma_1\,\sinh[2\pi  \delta(y+1)]\,\cos(2\pi x)\,,\\
& \hat{u}(x,y)=-\, \delta\gamma_1\,\cosh[2\pi \delta(y+1)]\,\sin(2\pi x)\,,
\end{align*} 
\end{linenomath*}
and the associated dynamic pressure is
\beq 
\hp(x,y)=-\delta \gamma_1 \Big\{[\hat c-\omega(1+y)]\cosh[2 \pi \delta (y+1)] + \f{\omega}{2\pi \delta}\sinh[2 \pi \delta (y+1)]\Big\}\sin(2\pi x)\,,
\label{edpl}
\eeq
where 
\begin{linenomath*}
\begin{equation}\label{A}
A=-\gamma_1 \delta\Big\{  [\hat{c}-\omega]\,\cosh(2\pi \delta) +\frac{\omega}{2\pi\delta}\,\sinh(2\pi \delta)\Big\}>0
\end{equation} 
\end{linenomath*}
(see the discussion in Appendix 1).

Some general properties of the dynamic pressure follow directly from \eqref{edpl}. Indeed, note that along the surface $y=0$ and along the flat bed $y=-1$, by \eqref{edpl} we have
\beq 
\hp(x,0)=A \,\sin(2\pi x)
\label{edpl-sur}
\eeq
and
\beq 
\hp(x,-1)=- \delta\gamma_1\hat{c} \,\sin(2\pi x)
\label{edpl-bed}
\eeq
respectively, with $A$ given by \eqref{A}. Thus the extrema of the restriction of the dynamic pressure to the surface and to the bed are at the points
\begin{linenomath*}
\begin{eqnarray}
\hat{p}(\pm 1/4,0) &=& \pm A\,,\label{lptop} \\
\hat{p}(\pm 1/4,-1) &=& \mp\delta\gamma_1\hat{c}=\pm A\,\frac{\hat{c}}{[\hat{c}-\omega]\,\cosh[2\pi \delta] + 
\frac{\omega}{2\pi \delta}\,\sinh[2\pi \delta]}\,.\label{lpbed}
\end{eqnarray}
\end{linenomath*}
In order to find the locations where $\hat{p}$ attains its maximum and minimum throughout the fluid domain, we also have to investigate the 
behaviour of $\hat{p}$ along the two open vertical segments $\{(\pm 1/4,y):\ -1 < y < 0\}$. Computing from \eqref{edpl} the derivatives of $\hp$: 
\begin{align}
	\hp_x(x,y)&=-2 \pi \delta \gamma_1 \Big\{ [\hc-\omega(1+y)]\cosh[2\pi \delta (y+1)]+\frac{\omega}{2\pi\delta}\sinh[\dpd(y+1)]\Big\}\cos(2 \pi x) \label{lpx} \\
	\hp_y(x,y)&= -2 \pi \delta^2 \gamma_1 [\hc-\ome(1+y)]\sinh[2 \pi \delta(y+1)]  \sin(2\pi x) \label{lpy}
\end{align}
we see that  in the absence of flow reversal the function $\hat{p}$ is strictly monotone along both segments, so that the extrema of $\hat{p}$ do not belong 
to these vertical segments and the largest and smallest among the four numbers listed in \eqref{lptop}-\eqref{lpbed} provide the extremal values  
and the location where they are attained within the fluid domain. On the other hand, the presence of a critical layer might alter this. To compare the magnitudes of the four values in \eqref{lptop}-\eqref{lpbed}, note that elementary calculations 
(see Appendix 2) show that
\begin{linenomath*}
\begin{equation}\label{clp3}
-1<\frac{\hat{c}}{[\hat{c}-\omega]\,\cosh[2\pi \delta] + 
\frac{\omega}{2\pi \delta}\,\sinh[2\pi \delta]}<1\,,
\end{equation}
\end{linenomath*}
unless
\begin{linenomath*}
\begin{equation}\label{bol}
\omega^2 \ge \frac{ [\cosh(2\pi \delta)+1]^2 \frac{\tanh(2\pi \delta)}{2\pi \delta}}{ 1 + [\cosh(2\pi \delta)-1] \frac{\tanh(2\pi \delta)}{2\pi \delta} 
- \cosh(2\pi \delta)\,\frac{\tanh^2(2\pi \delta)}{(2\pi \delta)^2}}\,.
\end{equation}
\end{linenomath*}

\subsection{Flow-reversal and critical layers}

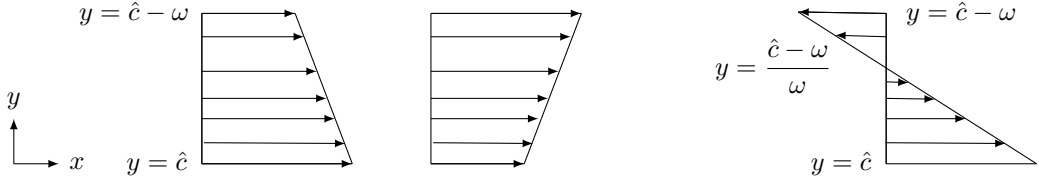
\begin{figure}
		\begin{minipage}{0.4 \textwidth}
			\centering
	\begin{tikzpicture}
\draw  [->,>=latex] (-2.5,0)--+(0.6,0) node[pos=1.4]{$x$}; 
\draw  [->,>=latex] (-2.5,0)--+(0.,0.6) node[pos=1.4]{$y$}; 
\draw [->,>=latex] (0,2) -- (1.24,2);
\draw [->,>=latex] (0,1.69) -- (1.36,1.69);
\draw [->,>=latex] (0,1.23) -- (1.53,1.23);
\draw [->,>=latex] (0,0.87) -- (1.66,0.87);
\draw [->,>=latex] (0,0.6) -- (1.77,0.6);
\draw [->,>=latex] (0.03,0.28) -- (1.91,0.27);
\draw [->,>=latex]  (0,0)-- (2,0);
\draw (0,2)-- (0,0);
\draw (0,0)-- (2,0);
\draw (0,0)-- (0,2);
\draw (1.24,2)-- (0,2);
\draw  (1.24,2)-- (2,0);
		\draw(-0.7,0) node{$y=-1$};
		\draw(-0.65,2) node{$y=0$};
		\draw(6,0) node{$y=-1$};
		\draw(6.4,2) node{$y=0$};
	\end{tikzpicture}
		\end{minipage}
		\begin{minipage}{0.2 \textwidth}
	\begin{tikzpicture}
\draw (0,2)-- (0,0);
\draw (0,0)-- (1.24,0);
\draw (2,2)-- (0,2);
\draw  (2,2)-- (1.24,0);
\draw [->,>=latex] (0,2) -- (2,2);
\draw [->,>=latex] (0,1.69) -- (1.86,1.69);
\draw [->,>=latex] (0,1.23) -- (1.7,1.23);
\draw [->,>=latex] (0,0.87) -- (1.56,0.87);
\draw [->,>=latex] (0,0.6) -- (1.43,0.6);
\draw [->,>=latex] (0.03,0.28) -- (1.36,0.27);
\draw [->,>=latex]  (0,0)-- (1.24,0);
	\end{tikzpicture}
		\end{minipage}
		\begin{minipage}{0.3 \textwidth}
			\centering
	\begin{tikzpicture}
\draw (0,2)-- (0,0);
\draw (0,0)-- (2,0);
\draw (0,0)-- (0,2);
\draw (-1.17,2.02)-- (0,2);
\draw (-1.17,2.02)-- (2,0);
\draw [->,>=latex] (0,2) -- (-1.17,2.02);
\draw [->,>=latex] (0,1.69) -- (-0.68,1.71);
\draw [->,>=latex] (0,1.09) -- (0.3,1.08);
\draw [->,>=latex] (0,0.87) -- (0.65,0.86);
\draw [->,>=latex] (0,0.6) -- (1.06,0.6);
\draw [->,>=latex] (-0.01,0.28) -- (1.59,0.27);
		\draw(0.55,2) node{$y=0$};
		\draw(-1.25,1.2) node{$y=\frac{\hc-\ome}{\ome}$};
		\draw(-0.66,0) node{$y=-1$};
	\end{tikzpicture}
		\end{minipage}
		\caption{Representation of the pure current $\hc-\omega(1+y)$, $y \in [-1,0]$.  Left: unidirectional flow for $\hc > \omega>0$. Middle: unidirectional flow for $\hc >0 >\omega$. Right: flow-reversal for $0< \hc < \omega$. }
		\label{fig:cw}
\end{figure}

Irrotational flow admits only uniform underlying currents, whose flow is obviously unidirectional. However, constant non-zero vorticity can bring about flow-reversal. 
Indeed, we can interpret the dispersion relation \eqref{eqfc2} as determining the difference between the wave speed $\hat c$ and the surface speed of the underlying current $(\omega(1+y),0)$. The travelling wave profile 
$h$ can move upstream or downstream relative to the the current $(\omega(1+y),0)$: the wave moves at speed $\hat c$, the underlying current has surface speed $\omega$, 
and the relative velocity (between surface current and wave) is $\hat c - \omega$. Consequently, the necessary and sufficient condition for flow reversal is that $y \mapsto [\hat{c}-\omega(1+y)]$ changes 
sign strictly on $[-1,0]$. Assuming $\hc \ge 0$, this is the case if
\begin{linenomath*}
\begin{equation}\label{frl0}
\hat{c}-\omega < 0\,,
\end{equation}
\end{linenomath*}
which requires $\omega > 0$. If \eqref{frl0} holds, then the critical level is  
\begin{linenomath*}
\begin{equation}\label{crl0}
y=\frac{\hat{c}-\omega}{\omega} \in (-1,0)\,.
\end{equation}
\end{linenomath*}
Due to \eqref{eqfc2}, the necessary and sufficient conditions for flow-reversal are 
\begin{linenomath*}
\begin{equation}\label{frl}
\omega >0  \quad\text{and}\quad 
-\,\frac{\tanh(2\pi \delta)}{4\pi  \delta} - \sqrt{\frac{\tanh^2(2\pi  \delta)}{(4\pi  \delta)^2} + 
\frac{\tanh(2\pi  \delta)}{2\pi  \delta \omega^2}} \in (-1, 0)\,,
\end{equation}
\end{linenomath*}
with the choice of a $-$ sign in the dispersion relation \eqref{eqfc2}. Fig. \ref{fig:cw} illustrates the underlying current pattern $\hc-\omega(1+y)$ with and without flow-reversal.

To further investigate the flow-reversal scenario it is convenient to have a closer look at the dispersion relation \eqref{eqfc2}.

\subsubsection{Stationary waves}
\label{sec:sw}

We note that $\hat{c}=0$ is possible in \eqref{eqfc2} but only when
\begin{linenomath*} 
\begin{equation}\label{block}
\omega^2=\frac{\tanh(2\pi\delta)}{2\pi\delta-\tanh(2\pi\delta)}\,,
\end{equation}
\end{linenomath*}
irrespective of the choice of sign $\pm$; this corresponds to special circumstances in which stationary waves are observed. This phenomenon does not occur for irrotational flow, being only possible for a given current of 
constant vorticity if the wavelength (encoded in $\delta$) of the waves reaches a specific value. Indeed, by computing the derivative, we see that the function $\delta 
\mapsto \frac{\tanh(2\pi \delta)}{2\pi\delta-\tanh(2\pi \delta)}$ is a strictly decreasing bijection from $(0,\infty)$ to $(0,\infty)$, 
so that the equation \eqref{block} has a unique positive root $\delta^\ast >0$, corresponding to the wavelength $L^\ast=d/\delta^\ast$. One can also consider the situation when swell originating from 
a distant storm interacts with a current of constant vorticity (for example, a tidal current, with positive constant vorticity
$\omega>0$ appropriate for the ebb and negative constant vorticity $\omega<0$ for the flood -- see the discussion in \cite{ck}). Due to \eqref{delta}, the swell wavelength $L$ fixes $\delta=d/L$, and \eqref{block} yields 
\begin{linenomath*} 
\begin{equation}\label{sw}
\omega= \pm \sqrt{\frac{\tanh(2\pi\delta)}{2\pi\delta-\tanh(2\pi\delta)}}\,.
\end{equation}
\end{linenomath*}
From the Taylor series expansion
\begin{linenomath*} 
$$\tanh(s)= s - \frac{s^3}{3} + \frac{2x^5}{15} - \frac{17 x^7}{315} + \dots$$
\end{linenomath*}
of the hyperbolic tangent function near the origin, we see that the solutions to \eqref{sw} approach $\pm \infty$ in the shallow water limit $\delta \to 0$. This shows that the ``wave blocking" phenomenon, in 
which stationary wave patterns are formed when wind-driven ocean waves encounter an opposing current that matches the wave speed (with $\hat{c}=0$ as the outcome of the wave.-current interaction), is not to be expected in the shallow water regime. 
However, since the function $s \mapsto \frac{\tanh(s)}{s- \tanh(s)}$ is a strictly decreasing bijection from $(0,\infty)$ onto itself, for any given $L$ the equation \eqref{sw} has a unique solution $\omega>0$. In the 
deep-water regime $\delta \to \infty$ of short waves in water of fixed depth $d$, given that $0.999 <  \tanh(s) <1$ for $s \ge 2\pi$, we can replace \eqref{sw} with
\begin{linenomath*} 
\begin{equation}\label{swd}
\omega \approx \pm \frac{1}{\sqrt{2\pi\delta-1}}= \pm \sqrt{\frac{L}{2\pi d -L}}\quad\text{for}\quad \delta=\frac{d}{L}> 1 \,.
\end{equation}
\end{linenomath*}
Since the function $L \mapsto \frac{L}{2\pi d -L}$ is increasing on $(0,2\pi d)$, we see that longer waves require stronger currents to generate stationary wave profiles by blocking. 
We refer to the 0:52-0:57 segment in the clip available at https://www.youtube.com/watch?v=yVW6XclyBak for a fascinating video of wave blocking by a tidal current in the 100 m deep Seymour Narrows in British Columbia; in this context a short wavelength 
$L=5$ m and a surface current speed of  about 3 m$\,$s$^{-1}$ yield a solution to \eqref{sw} with $\omega \approx 0.09$ (corresponding to $\Omega=\omega \sqrt{g/d} \approx 0.03$ s$^{-1}$).

\subsubsection{Waves propagating to the right}
\label{sec:wpr}

Thinking of the wave-current interaction 
coming about as small-amplitude waves of specific wavelengths that start to appear at the surface of a pre-existing flow 
with constant vorticity $\omega$, for wave propagation to the right, with speed
\begin{linenomath*} 
\begin{equation}\label{ws}
c= \hat{c}\,\sqrt{gd} >0
\end{equation}
\end{linenomath*}
we distinguish the following cases:
\begin{itemize}
	\item $\ome=0 $ (irrotational flow): the dispersion relation \eqref{eqfc2}  and \eqref{ws} yield 
\begin{linenomath*} 
\begin{equation}\label{sir}
c=\sqrt{gd\,\frac{\tanh(2\pi \delta)}{2\pi\delta}} >0
\end{equation}
\end{linenomath*}

\item for $\ome>0$ (favorable current) there are two possibilities:
		\begin{itemize}
			\item we have \textit{fast travelling waves} with speed
				\beq 
				c=\sqrt{gd} \left( \ome \left( 1-\f{\tanh(\dpd)}{4\pi \delta} \right) +\sqrt{ \left(\f{\ome \tanh(\dpd)}{4\pi \delta} \right)^2+\f{\tanh{\dpd}}{\dpd}}  \right)>0
				\label{fw}
				\eeq
			\item for small wavelengths $L$, more precisely, for $L<L^*$ where $L^\ast$ corresponds to the solution $\delta=d/L$ of \eqref{sw} with fixed vorticity $\omega$ and depth $d$, so that
				\beq 
				\ome^2 > \f{\tanh(2\pi \delta)}{\dpd-\tanh(\dpd)} \,,
				\label{excl}
				\eeq
				we also have \textit{slow waves} propagating with speed
				\beq 
				c=\sqrt{gd} \left( \ome \left( 1-\f{\tanh(\dpd)}{4\pi \delta} \right) -\sqrt{ \left(\f{\ome \tanh(\dpd)}{4\pi \delta} \right)^2+\f{\tanh{\dpd}}{\dpd} } \right)>0 \,.
				\label{sws}
				\eeq
				
		\end{itemize}
			\item for $\ome <0$ (adverse current), waves propagating to the right only exist if the wavelength $L$ is large enough, more precisely, if $L>L^*$  where $L^\ast$ 
			corresponds to the solution $\delta=d/L$ of \eqref{sw} with fixed vorticity $\omega$ and depth $d$, so that
				\beq 
				\ome^2 < \f{\tanh(2\pi \delta)}{\dpd-\tanh(\dpd)} \,,
\label{excl2}
				\eeq
in which case the speed of propagation is
\begin{linenomath*} 
\begin{equation}\label{wa}
c=\sqrt{gd}\,\bigg\{\omega \Big(1-\frac{\tanh(2\pi \delta)}{4\pi\delta}\Big) + \sqrt{\frac{\omega^2\tanh^2(2\pi \delta)}{(4\pi\delta)^2} 
+\frac{\tanh(2\pi \delta)}{2\pi\delta} } \ \bigg\} >0\,.
\end{equation}
\end{linenomath*}
The relation \eqref{excl2} can be interpreted as a constraint on the strength of the adverse current -- a too strong current impedes wave propagation (see e.g. the segment 3:15-3:27 of the video 
clip available at https://www.youtube.com/watch?v=yVW6XclyBak). 
\end{itemize}

\section{Extrema of the dynamic pressure}
\label{sec:dyn}

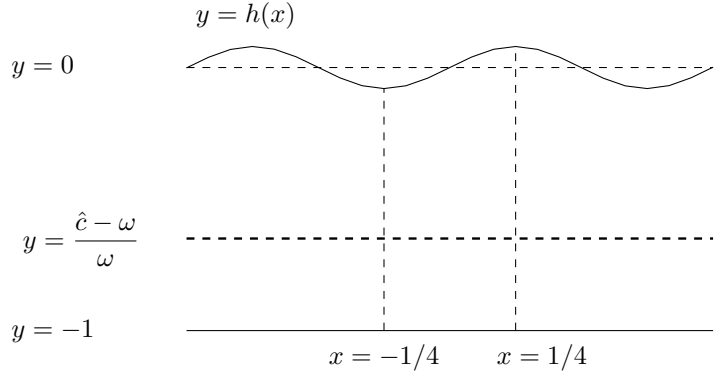
\begin{figure}[htbp]
\centering
\begin{tikzpicture}[scale=3.5]
\draw(-1,-1)--(1,-1);
\draw[dashed](0.25,-1)--+(0,1.08);
\draw[dashed](-0.25,-1)--+(0,0.92);
\draw [domain=-1:1] plot (\x,{sin(6.3*\x r)*0.08});
\draw[dashed](-1,0)--(1,0);
\draw(-1.7,-1) node[right]{$y=-1$};
\draw(-1.7,0) node[right]{$y=0$};
\draw(-1,0.2) node[right]{$y=h(x)$};
	\draw[dashed,line width=0.9pt](-1,-0.65)--+(2,0.) node[pos=-0.2]{$y=\f{\hc-\ome}{\ome}$};
	\draw(-0.25,-1.1) node{$x=-1/4$};
	\draw(0.35,-1.1) node{$x=1/4$};
\end{tikzpicture}
	\caption{ In a non-dimensional reference frame moving at the wave speed $\hat{c}$, the linear waves 
  $h(x)=A\sin(2\pi x)$ with principal period $1$ are steady sinusoidal oscillations of the 
  flat free surface $y=0$, with the wave crest/trough at $x=\pm\tfrac{1}{4}$. The necessary and sufficient condition for the existence of 
  a critical line $y=\frac{\hat{c}-\omega}{\omega}$, where the reversal of the underlying mean flow occurs, is \eqref{frl}. }
\label{fig:reduce}
\end{figure}

We now determine the location of the extrema of the dynamic pressure, defined in \eqref{pn}, throughout the fluid domain. For symmetry reasons we can restrict the study to the rectangle
\begin{linenomath*}
\begin{equation}\label{perb}
\{(x,y):\ -1 \le y \le 0\,,\quad -1/4 \le x \le 1/4 \}\,,
\end{equation}
\end{linenomath*}
which is the  fluid region between a consecutive crest and trough (see Figure \ref{fig:reduce}), the wave crest/trough being located at $x=\pm 1/4$.   

\subsection{Irrotational flow}

\begin{figure}[htbp]
\centering
\begin{tikzpicture}[scale=9]
\draw[smooth,samples=100,domain=1.25:1.75] plot(\x,0);
\draw (1.75,-0.42) -- (1.75,0);
\draw (1.25,0)-- (1.25,-0.42);
\draw (1.25,-0.42)-- (1.75,-0.42);
\draw [dash pattern=on 2pt off 2pt,cyan,thick] (1.5,0)-- (1.5,-0.42);
			\draw(1.15,0) node{min$(\hat{p})$};
			\draw(1.85,0) node{max$(\hat{p})$};
				\fill(1.75,0) circle(0.02);
				\fill(1.25,0) circle(0.01);
			\draw(1.4,-0.15) node{$p_x>0$};
			\draw(1.4,-0.25) node{$p_y<0$};
			\draw(1.6,-0.15) node{$p_x>0$};
			\draw(1.6,-0.25) node{$p_y>0$};
			\draw(1.25,-0.45) node{$x=-1/4$};
			\draw(1.75,-0.45) node{$x=1/4$};
			\draw[cyan](1.5,-0.45) node{$x=0$};
\end{tikzpicture}
\caption{The monotonicity of the dynamic pressure in a periodicity box beneath an irrotational wave (linear theory), in 
  accordance to the relations in \eqref{derpzx}-\eqref{derpzyl}. The maximum/minimum of the dynamic pressure $\hat{p}$ are attained at the wave 
  crest/trough.}
\label{fig:pmain}
\end{figure}
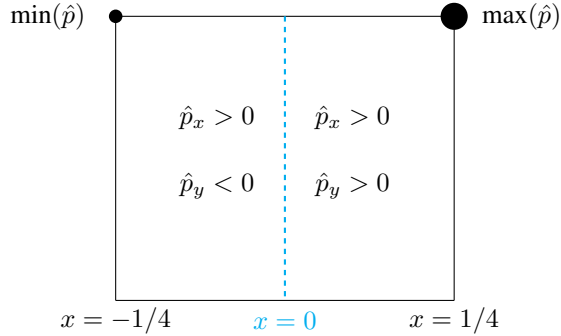

For $\omega=0$, by \eqref{sir}, the nondimensional speed of propagation to the right is 
\begin{linenomath*} 
$$\hat{c} = \frac{c}{\sqrt{gd}} =\sqrt{ \frac{\tanh(2\pi  \delta)}{2\pi  \delta}} > 0\,.$$
\label{eq:cirrot}
\end{linenomath*}
Combining \eqref{A} with \eqref{edpl}, the dynamic pressure is given by
\begin{linenomath*} 
$$\hat{p}(x,y)= \frac{A}{\cosh(2\pi\delta)}\,\cosh[2\pi \delta(y+1)] \,\sin(2\pi x),\qquad x \in {\mathbb R}\,,\ -1 \le y \le 0\,,$$
\end{linenomath*}
with $A>0$, so that 
\begin{linenomath*} 
\begin{eqnarray}
&& \hat{p}_x(x,y) > 0 \ \hbox{for} \  (x,y) \in (-\tfrac{1}{4},\tfrac{1}{4}) \times [-1,0] \ \text{and}\ \hat{p}_x=0\ \hbox{on}\ x=\pm \tfrac{1}{4}\,,\label{derpzx}\\
&& \hat{p}_y(x,y) > 0 \ \hbox{for} \  (x,y) \in (0,\tfrac{1}{4}] \times (-1,0] \ \text{and}\ \hat{p}_y=0\ \hbox{on}\ y=-1\,,\label{derpzyr}\\
&& \hat{p}_y(x,y) < 0 \ \hbox{for} \  (x,y) \in [-\tfrac{1}{4},0) \times (-1,0]\ \text{and}\ \hat{p}_y=0\ \hbox{on}\ x=0\,,\label{derpzyl}
\end{eqnarray}
 \end{linenomath*}
and therefore the overall maximum/minimum of $\hat{p}$ is at the wave crest/trough; see Figure \ref{fig:pmain}. Note that this location of the extrema holds also for nonlinear waves (see 
\cite{c-pf}), in which case the proof relies not on the explicit expression, which is not available, but on maximum principles.

\subsection{Positive vorticity without flow-reversal}
\label{sec:cas2dyn}

Relation \eqref{frl} shows that the absence of flow-reversal for right-propagating waves occurs only for the choice of a $+$ sign in the dispersion relation \eqref{eqfc2} 
since the choice of the $-$ sign in the dispersion relation \eqref{eqfc2} would require
\begin{linenomath*}
$$-\,\frac{\tanh(2\pi \delta)}{4\pi  \delta} - \sqrt{\frac{\tanh^2(2\pi  \delta)}{(4\pi  \delta)^2} + 
\frac{\tanh(2\pi  \delta)}{2\pi  \delta \omega^2}} \le -1\,,$$
\end{linenomath*}
a situation that is only possible if 
\begin{linenomath*}
$$0 < \omega \le \sqrt{\frac{\tanh(2\pi \delta)}{2\pi \delta- \tanh(2\pi \delta)} }\,,$$
\end{linenomath*}
which, by \eqref{excl}, forces the waves to propagate to the left.

For the choice of the $+$ sign in dispersion relation \eqref{eqfc2} the discussion in the previous subsection shows that
\begin{linenomath*}
\begin{equation}\label{posnfr}
0 < \frac{\hat{c}}{[\hat{c}-\omega]\,\cosh(2\pi \delta) + \frac{\omega}{2\pi\delta}\,\sinh(2\pi\delta)} <1\,,
\end{equation}
\end{linenomath*}
so, in particular, \eqref{clp4} is validated. 
Consequently \eqref{lptop}-\eqref{lpbed} ensure that the overall maximum/minimum of $\hat{p}$ is attained at the wave crest/trough. 
From \eqref{posnfr} and \eqref{A} we deduce that $\gamma_1<0$. Since $\hat{c}>\hat{c}-\omega>0$ ensures that $\hat{c}-\omega(1+y) > 0$  for all $y \in [-1,0]$, from \eqref{lpx} (respectively \eqref{lpy}) we infer that \eqref{derpzx}-\eqref{derpzyr}-\eqref{derpzyl} hold also in this setting; see Figure \ref{fig:pmain} for an overall illustration.

\subsection{Negative vorticity}
\label{sec:cas3dyn}

In this setting, the choice of a $-$ sign in the dispersion relation \eqref{eqfc2} entails $\hat{c}<0$, so stationary waves or waves propagating to the right occur only for the choice of a 
$+$ sign in the dispersion relation \eqref{eqfc2}. Flow reversal does not occur (recall \eqref{frl}) and we have two possible scenarios:
\begin{itemize} 
\item The vorticity satisfies \eqref{bol}. Let us note that the function 
\begin{linenomath*}
$$y \mapsto \hat{c}-\omega(y+1) + \frac{\omega}{2\pi \delta}\,\tanh[2\pi \delta(y+1)]$$
\end{linenomath*}
equals $\hat{c}>0$ for $y=-1$ and is 
strictly increasing on $[-1,0]$ because its derivative equals $-\omega \tanh^2[2\pi \delta(y+1)]>0$. Consequently 
\begin{linenomath*}
\begin{equation}\label{y+n}
\hat{c}-\omega(y+1) + \frac{\omega}{2\pi \delta}\,\tanh[2\pi \delta(y+1)] >0\,,\qquad y \in [-1,0]\,,
\end{equation}
\end{linenomath*}
and \eqref{A} ensures $\gamma_1 < 0$. From \eqref{lpx} we now deduce that
\begin{linenomath*} 
\begin{equation}\label{derpxn}
\hat{p}_x(x,y) \geq 0 \ \hbox{for} \  x \in [-1/4,1/4]\,,\ y \in [-1,0]\,,\\
\end{equation}
 \end{linenomath*} 
while \eqref{lpy} yields
\begin{linenomath*} 
\begin{equation}\label{derpyn}
\begin{cases}
& \hat{p}_y(x,y) > 0 \ \hbox{for} \  (x,y) \in (0,1/4] \times (-1,0] \ \text{and}\ \hat{p}_y=0\ \hbox{on}\ y=-1\,,\\
& \hat{p}_y(x,y) < 0 \ \hbox{for} \  (x,y) \in [-1/4,0) \times (-1,0]\ \text{and}\ \hat{p}_y=0\ \hbox{on}\ x=0\,,
\end{cases}
\end{equation}
 \end{linenomath*} 
since $\hat{c}-\omega(y+1) >0$ for $y \in [-1,0]$ due to \eqref{y+n}. The lack of flow reversal and 
a comparison of the four values listed in \eqref{lptop}-\eqref{lpbed} ensures that the maxima/minima of the dynamic pressure in the fluid domain are attained at the crest/trough. 
The situation is that depicted in Figure \ref{fig:pmain}.

\item The vorticity does not satisfy \eqref{bol}. In this case, the discussion in Section 3.2 shows 
the validity of \eqref{clp3}. From \eqref{lptop}-\eqref{lpbed} we  
deduce that the overall maximum/minimum of $\hat{p}$ is at the wave trough/crest.  Furthermore, $\omega < 0$ yields 
$\gamma_1<0$ from \eqref{A}, since $\tanh(2\pi\delta)<2\pi\delta$ holds for all $\delta >0$. From \eqref{lpy} we deduce that
\begin{linenomath*} 
\begin{equation}\label{derpy}
\begin{cases}
& \hat{p}_y(x,y) > 0 \ \hbox{for} \  (x,y) \in (0,1/4] \times (-1,0] \ \text{and}\ \hat{p}_y=0\ \hbox{on}\ y=-1\,,\\
& \hat{p}_y(x,y) < 0 \ \hbox{for} \  (x,y) \in [-1/4,0) \times (-1,0]\ \text{and}\ \hat{p}_y=0\ \hbox{on}\ x=0\,,
\end{cases}
\end{equation}
 \end{linenomath*} 
On the other hand, $\hat{c} > 0$ in \eqref{lpx} yields $\hat{p}_x=0$ for $x=\pm  1/4$ and
\begin{linenomath*} 
\begin{equation}\label{derpx}
\hat{p}_x(x,-1) \geq 0 \ \hbox{for} \  x \in [-1/4,1/4]\,. 
\end{equation}
 \end{linenomath*} 
We also have that 
\begin{linenomath*}
\begin{eqnarray*}
&& [\hat{c}-\omega(1+y)]\,\cosh[2\pi\delta(y+1)] + \frac{\omega}{2\pi\delta}\,\sinh[2\pi\delta(y+1)]  \\
&&\quad > -\omega\Big\{ (1+y)\cosh[2\pi\delta(y+1)] - \frac{\sinh[2\pi\delta(y+1)]}{2\pi\delta} \Big\} >0\,,\qquad y \in (-1,0]\,,
\end{eqnarray*}
\end{linenomath*}
since the function 
\begin{linenomath*}
$$y \mapsto (1+y)\cosh[2\pi\delta(y+1)] - \frac{\sinh[2\pi\delta(y+1)]}{2\pi\delta}$$
\end{linenomath*}
has a positive derivative on $(-1,0)$ and vanishes at $y=-1$. Using \eqref{lpx}, we deduce that for $\hc>0$ we have
\begin{linenomath*} 
\begin{equation}\label{derpx}
\hat{p}_x(x,y) > 0 \ \hbox{for} \  (x,y) \in (-1/4,1/4) \times [-1,0] \ \text{and}\ \hat{p}_x=0\ \hbox{on}\ x=\pm 1/4.
\end{equation}
\end{linenomath*} 
and the situation illustrated in Figure \ref{fig:pmain} remains valid. Note that stationary waves with $\hc=0$ may occur, if \eqref{sw} holds with the $-$ sign. This is a limiting case 
of the situation illustrated in Figure \ref{fig:pmain}, the only difference being that in this special setting the dynamic pressure is constant along the flat bed since $\hat{p}_x(x,-1)=0$. 
\end{itemize}

\subsection{Positive vorticity with flow-reversal}
\label{sec:cas4dyn}

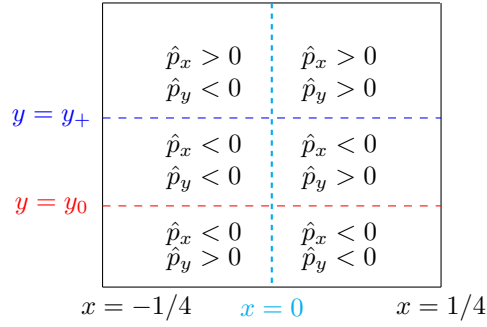
\begin{figure}[htbp]
	\centering
\begin{tikzpicture}[scale=9]
\draw[smooth,samples=100,domain=1.25:1.75] plot(\x,0);
\draw (1.75,-0.42) -- (1.75,0);
\draw (1.25,0)-- (1.25,-0.42);
\draw (1.25,-0.42)-- (1.75,-0.42);
\draw [dash pattern=on 2pt off 2pt,cyan,thick] (1.5,0)-- (1.5,-0.42);
				
				\draw[blue,dashed] (1.25,-0.17)--(1.75,-0.17) node[pos=-0.15]{$y=y_+$};
				\draw[red,dashed] (1.25,-0.3)--(1.75,-0.3) node[pos=-0.15]{$y=y_0$};
		\draw(1.4,-0.08) node{$\hp_x>0$};
			\draw(1.4,-0.13) node{$\hp_y<0$};
				\draw(1.6,-0.08) node{$\hp_x>0$};
			\draw(1.6,-0.13) node{$\hp_y>0$};

		\draw(1.4,-0.21) node{$\hp_x<0$};
				\draw(1.4,-0.26) node{$\hp_y<0$};
				\draw(1.6,-0.21) node{$\hp_x<0$};
			\draw(1.6,-0.26) node{$\hp_y>0$};

		\draw(1.4,-0.34) node{$\hp_x<0$};
				\draw(1.4,-0.38) node{$\hp_y>0$};
				\draw(1.6,-0.34) node{$\hp_x<0$};
			\draw(1.6,-0.38) node{$\hp_y<0$};
			\draw(1.3,-0.45) node{$x=-1/4$};
			\draw(1.75,-0.45) node{$x=1/4$};
			\draw[cyan](1.5,-0.45) node{$x=0$};

\end{tikzpicture}
\caption{The monotonicity of the dynamic pressure $\hat{p}$ in the periodicity box with surface $y=0$ and bed $y=-1$ for vorticity 
  $\omega>0$ and flow reversal across the critical line $y=y_0 \in (-1,0)$. According to \eqref{derpy}-\eqref{derpx} $\hat{p}_x$ changes 
  sign across the level $y=y_+ \in (y_0,0)$, where $y_+ \in (y_0,-1)$ is the unique solution of equation \eqref{y+}, while $\hat{p}_y$ changes sign across the critical line $y=y_0$ 
  and across $x=0$.}
  \label{fig:reversalvar}
\end{figure}

Relation \eqref{frl}, which is equivalent to \eqref{excl} and to 
\begin{linenomath*} 
\begin{equation}\label{frpv}
1-\f{\tanh(\dpd)}{4\pi \delta}  > \sqrt{ \left(\f{ \tanh(\dpd)}{4\pi \delta} \right)^2+\f{\tanh{\dpd}}{\dpd \omega^2} }\,,
\end{equation}
\end{linenomath*} 
with $\omega>0$, together with the choice of a $-$ sign in the dispersion relation \eqref{eqfc2} so that $0<\hat{c}<\omega$, specifies when flow-reversal occurs. 
In this setting we only have slow waves propagating to the right with speed \eqref{sws} and small wavelengths $L<L^\ast$. 
The dispersion relation \eqref{eqfc2} with a $-$ sign yields
\begin{linenomath*} 
\begin{equation}\label{fri1}
[\hat{c}-\omega]\,\cosh(2\pi\delta) + \frac{\omega\sinh(2\pi\delta)}{2\pi\delta}  =\frac{\omega \sinh(2\pi\delta)}{4\pi\delta}
- \sqrt{\frac{\omega^2\sinh^2(2\pi\delta)}{(4\pi\delta)^2}+\frac{\sinh(4\pi\delta)}{4\pi\delta}} <0\,,
\end{equation}
\end{linenomath*} 
and from \eqref{A} we obtain $\gamma_1>0$. Relation \eqref{lpy} shows that $\hat{p}_y(x,y)=0$ in the periodicity box \eqref{perb} precisely along $x=0$, along the flat bed $y=-1$ and 
along the critical line $y=y_0$, with $y_0=\frac{\hat{c}-\omega}{\omega}$ determined in \eqref{crl0}. From \eqref{lpy} we can also determine the sign of $\hat{p}_y(x,0)$, inferring that in the interior of the periodicity box \eqref{perb} 
we have $\hat{p}_y(x,y)<0$ if $x<0$ and $y>y_0$, while $\hat{p}_y(x,y)>0$ if $x>0$ and $y>y_0$. 
On the other hand, since $y<y_0$ ensures $\hat{c}-\omega(1+y)>0$, we have that $\hat{p}_y(x,y)>0$ if $x<0$ and 
$y<y_0$, while $\hat{p}_y(x,y)<0$ if $x>0$ and $y<y_0$ in the periodicity box \eqref{perb}. As for the sign of $\hat{p_x}$ in 
the interior of the periodicity box \eqref{perb}, \eqref{lpx} shows that $\hat{p}_x(x,y)>0$ for $y>y_+$ and $\hat{p}_x(x,y)<0$ for $y<y_+$, 
where $y_+ \in (y_0,0)$ is the unique solution in $[-1,0]$ of the equation
\begin{linenomath*} 
\begin{equation}\label{y+}
\omega(1+y) - \hat{c}-\frac{\omega}{2\pi\delta}\,\tanh[2\pi\delta(1+y)]=0\,,
\end{equation}
\end{linenomath*} 
whose right side is positive at $y=0$, in view of \eqref{fri1}, negative at $y=y_0>-1$ and has derivative 
$\omega\tanh^2[2\pi\delta(1+y)] >0$ on $(-1,0)$. We summarize these monotonicity properties in Figure \ref{fig:reversalvar}. They ensure that 
in the periodicity box \eqref{perb} there are only six possible locations for the maximum of $\hat{p}$, at $(1/4,0)$ or at $(- 1/4,y_0)$, while the minimum 
is attained at $(-1/4,0)$ or at $(1/4,y_0)$.

Note that \eqref{edpl} yields
\begin{linenomath*} 
\begin{eqnarray*}
&& \hat{p}(\pm \tfrac{1}{4},0) = \mp \delta\gamma_1\Big\{ [\hat{c}-\omega]\,\cosh(2\pi\delta) + \frac{\omega}{2\pi\delta}\,\sinh(2\pi\delta)\Big\}\,,\\
&& \hat{p}(\pm \tfrac{1}{4},y_0) = \mp  \delta\gamma_1\,\frac{\omega}{2\pi\delta}\,\sinh\Big(\frac{2\pi\delta \hat{c}}{\omega}\Big)\,,\\
&& \hat{p}(\pm \tfrac{1}{4},-1) = \mp \delta\gamma_1 \hat{c}\,,
\end{eqnarray*}
\end{linenomath*} 
with $\gamma_1>0$ and $\hat{c} \in (0,\omega)$. Using \eqref{fri1} and the fact that $\frac{\sinh(s)}{s}>1$ for $s>0$ (in particular for $s=2\pi \delta \hc/\ome$), we get
\begin{linenomath*} 
\begin{equation}\label{locep}
[\hat{c}-\omega]\,\cosh(2\pi\delta) + \frac{\omega}{2\pi\delta}\,\sinh(2\pi\delta) < 0 < \hat{c} < 
\frac{\omega}{2\pi\delta}\,\sinh\Big(\frac{2\pi\delta\hat{c}}{\omega}\Big)\,.
\end{equation}
\end{linenomath*} 
This confirms that the location of the maximum and minimum of $\hat{p}$ is either at the wave crest/trough or on the critical line 
$y=y_0$, just below the wave crest/trough; in this context it is useful to observe that, by \eqref{edpl}, the function $\hat{p}(x,y)$ is 
odd in $x$.  While the determination of the wavelength range for which the right side of \eqref{locep} exceeds the absolute value of the 
left side seems intractable by a direct approach, this can be done indirectly by relying on the monotonicity properties of $\hat{p}$.

Before discussing the possible configurations, let us note that 
\begin{linenomath*}
$$\frac{(2\pi \delta) [\cosh(2\pi \delta)+1]^2}{2\pi\delta+\sinh(2\pi \delta)} > \frac{(2\pi \delta)+  (2\pi \delta)[\cosh(2\pi \delta)]^2}{2\pi\delta+\sinh(2\pi \delta)} >1$$
\end{linenomath*} 
since the function $s \mapsto s [\cosh(s)]^2 - \sinh(s)$ vanishes in the limit $s \to 0$ and has a strictly positive derivative for $s>0$.

We can now proceed with a detailed analysis of the possible flow scenarios.
\begin{itemize}

\item If $\omega>0$ is such that 
\begin{linenomath*}
\begin{equation}\label{cl2} 
\frac{\tanh(2\pi \delta)}{2\pi\delta-\tanh(2\pi \delta)}\,\frac{(2\pi \delta) [\cosh(2\pi \delta)+1]^2}{2\pi\delta+\sinh(2\pi \delta)} \le \omega^2\,,
\end{equation}
\end{linenomath*}
then the discussion in Section 3.2, with \eqref{bol} valid, in combination with \eqref{lptop}-\eqref{lpbed}, shows that the extrema of 
$\hat{p}$ on the top and bottom boundary of the periodicity box \eqref{perb} 
(that is, along $y=0$ and $y=-1$) are attained at the lower corners: the minimum at $(1/4,-1)$ and the maximum at $(-1/4,-1)$. From 
the inequality \eqref{locep} we now deduce that the extrema of $\hat{p}_y$ are attained 
along the critical layer, the minimum at $(1/4,y_0)$ and the maximum at $(-1/4,y_0)$; see Figure \ref{fig:reversal1}. 

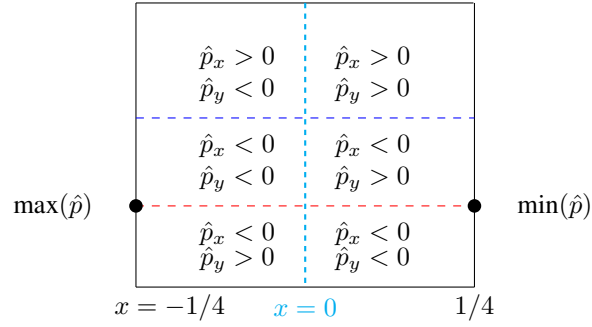
\begin{figure}[htbp]
\centering
\begin{tikzpicture}[scale=9]
\draw[smooth,samples=100,domain=1.25:1.75] plot(\x,0);
\draw (1.75,-0.42) -- (1.75,0);
\draw (1.25,0)-- (1.25,-0.42);
\draw (1.25,-0.42)-- (1.75,-0.42);
\draw [dash pattern=on 2pt off 2pt,cyan,thick] (1.5,0)-- (1.5,-0.42);
				\draw[red,dashed] (1.25,-0.3)--(1.75,-0.3);
				\fill(1.75,-0.3) circle(0.01);
				\fill(1.25,-0.3) circle(0.01);
				\draw(1.2,-0.3) node[left]{max$(\hat{p})$};
				\draw(1.8,-0.3) node[right]{min$(\hat{p})$};

				\draw[blue,dashed] (1.25,-0.17)--(1.75,-0.17);
		\draw(1.4,-0.08) node{$\hp_x>0$};
			\draw(1.4,-0.13) node{$\hp_y<0$};
				\draw(1.6,-0.08) node{$\hp_x>0$};
			\draw(1.6,-0.13) node{$\hp_y>0$};

		\draw(1.4,-0.21) node{$\hp_x<0$};
				\draw(1.4,-0.26) node{$\hp_y<0$};
				\draw(1.6,-0.21) node{$\hp_x<0$};
			\draw(1.6,-0.26) node{$\hp_y>0$};

		\draw(1.4,-0.34) node{$\hp_x<0$};
				\draw(1.4,-0.38) node{$\hp_y>0$};
				\draw(1.6,-0.34) node{$\hp_x<0$};
			\draw(1.6,-0.38) node{$\hp_y<0$};
			\draw(1.3,-0.45) node{$x=-1/4$};
			\draw(1.75,-0.45) node{$1/4$};
			\draw[cyan](1.5,-0.45) node{$x=0$};
\end{tikzpicture}
\caption{The maxima/minima of the dynamic pressure $\hat{p}$ for right-propagating waves with constant vorticity $\omega>0$ and flow-reversal occur along the critical line $y=y_0$, below the trough and crest, respectively.}
\label{fig:reversal1}
\end{figure}

\item If $\omega$ is such that 
\begin{linenomath*}
\begin{equation}\label{cl1}
\frac{\tanh(2\pi \delta)}{2\pi\delta-\tanh(2\pi \delta)} < \omega^2 < 
\frac{\tanh(2\pi \delta)}{2\pi\delta-\tanh(2\pi \delta)}\,\frac{(2\pi \delta) [\cosh(2\pi \delta)+1]^2}{2\pi\delta+\sinh(2\pi \delta)}\,,
\end{equation}
\end{linenomath*}
with the expression on the right equal to that on the right of \eqref{bol}, then the discussion in Section 3.2 shows that \eqref{clp3} holds, which ensures 
that the extrema of $\hat{p}$ on the top and bottom boundary of the periodicity box \eqref{perb} 
are attained at the upper corners: the minimum at $(-1/4,0)$ and the maximum at $(1/4,0)$. Since in this case 
\begin{linenomath*}
$$0 > [\hat{c}-\omega]\,\cosh(2\pi\delta) + \frac{\omega}{2\pi\delta}\,\sinh(2\pi\delta) > -\hat{c}\,,$$
\end{linenomath*}
from \eqref{locep} and from the sign of $\hat{p}_y$ in the regions delimited by 
the critical layer $y=y_0$ and by the vertical segments $x=0$ and $x=\pm 1/4$ 
enables us to conclude that the maximum of the dynamic pressure is attained either at the wave crest $(1/4,0)$ or at 
the point $(-1/4,y_0)$ on the critical level beneath the wave trough, while the minimum is attained either at the wave trough 
$(-1/4,0)$ or at the point $(1/4,y_0)$ on the critical level beneath the wave crest. Recalling \eqref{crl0} and the fact that $\gamma_1>0$, from \eqref{edpl} we compute
\begin{linenomath*}
\begin{align}
&0 < \hat{p}(1/4,0) = -\delta\gamma_1 \Big\{ [\hat{c}-\omega]\,\cosh(2\pi\delta) + \frac{\omega}{2\pi\delta}\,\sinh(2\pi\delta) \Big\} =- \hat{p}(-1/4,0)\,,\quad \label{ctc} \\
&0< \hat{p}(-1/4,y_0) = \delta\gamma_1 \, \frac{\omega}{2\pi\delta}\,\sinh[2\pi\delta(y_0+1)] =- \hat{p}(1/4,y_0)\,.\quad \label{clc}
\end{align} 
\end{linenomath*}
This shows that either the maximum of the dynamic pressure is at the wave crest $(1/4,0)$ and the minimum is at the wave trough $(-1/4,0)$, or both extrema 
are attained on the critical level $y=y_0$, beneath the trough and the crest, respectively. Due to \eqref{fri1}, we see that  $\hat{p}(1/4,0) > \hat{p}(-1/4,y_0)$ is equivalent to
\begin{linenomath*}
\begin{equation}\label{dec}
 \pi\delta\,\sinh(4\pi\delta)>  \omega^2\,\sinh[2\pi\delta(y_0+1)]  \Big\{\sinh(2\pi\delta) + \sinh[2\pi\delta(y_0+1)]  \Big\}\,.
\end{equation}
\end{linenomath*}

To gain further insight we have to investigate in more detail the possible locations of the critical level $y=y_0$ relative to the free surface $y=0$ and to the flat bed $y=-1$. 
For this, we use \eqref{fri1} and \eqref{crl0} to obtain 
\begin{linenomath*}
\begin{equation}\label{lsp}
y_0=\frac{\hat{c}-\omega}{\omega}= -\frac{\tanh(2\pi\delta)}{4\pi\delta}\,\Big\{ 1 + \sqrt{ 1 + \frac{8\pi\delta}{\omega^2\tanh(2\pi\delta)}} \Big\}\,.
\end{equation}
\end{linenomath*}
For a fixed wavelength (that is, with $\delta$ fixed), we infer from \eqref{lsp} that the height of the level set above the flat bed increases as the vorticity $\omega>0$ 
increases. For the lower bound on $\omega>0$ in \eqref{cl1} we obtain $y_0=-1$ in \eqref{lsp}, while for the upper bound \eqref{lsp} we obtain 
$y_0=- \frac{2\pi\delta + \sinh(2\pi\delta)}{2\pi\delta + 2\pi\delta\cosh(2\pi\delta)}$. Note that the function $s \mapsto \frac{s+\sinh(s)}{s + s \cosh(s)}$ being strictly decreasing 
on $(0,\infty)$ and with limit $1$ as $s \to 0$ and limit zero as $s \to \infty$. Thus, for $\omega>0$ subject to the constraint \eqref{cl1}, we may write \eqref{dec} in the form
\begin{linenomath*}
\begin{equation}\label{dec2}
\omega^2 < \frac{\pi\delta\,\sinh(4\pi\delta)  }{\sinh[2\pi\delta(y_0+1)]  \Big\{\sinh(2\pi\delta) + \sinh[2\pi\delta(y_0+1)]  \Big\}}\,,
\end{equation}
\end{linenomath*}
and the above considerations show that as $\omega>0$ approaches the lower bound in \eqref{cl1} the right side of \eqref{dec2} approaches $+\infty$ (and thus the 
maxima/minima of the dynamic pressure are at the wave crest/trough, while as $\omega>0$ approaches the upper bound in \eqref{cl1} the discussion of case \eqref{cl2} shows that 
the extrema of the dynamic pressure are along the critical level $y=y_0$, as depicted in Fig. \ref{fig:reversal1}.
\end{itemize}

\begin{figure}[htbp]
	\centering
	\includegraphics[scale=0.6]{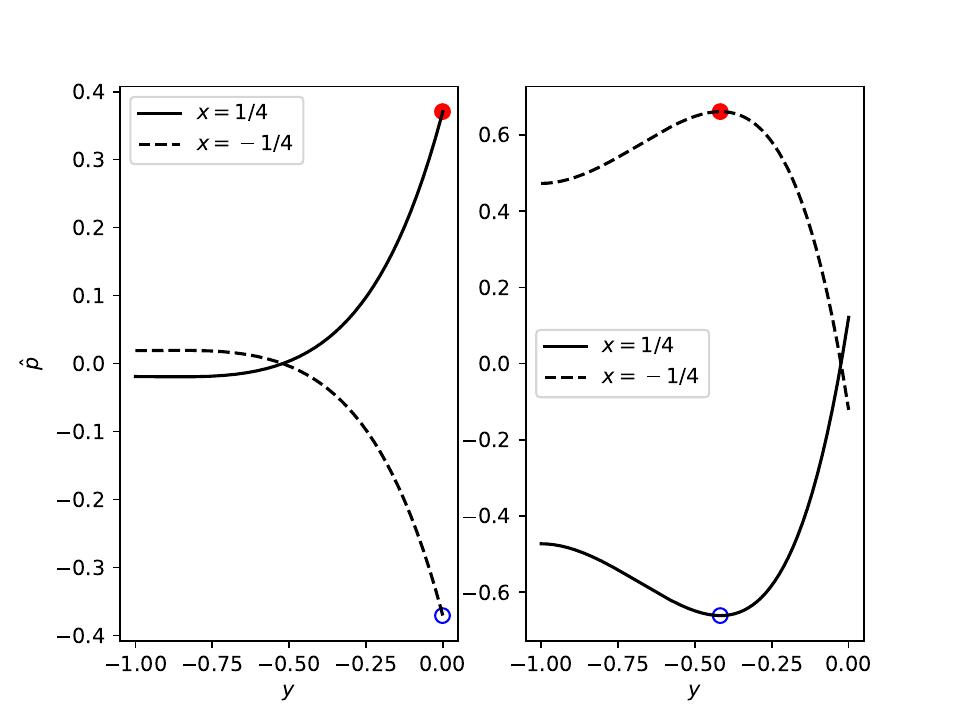}
	\caption{Two graphs of the dynamic pressure along the crest/trough lines $x=\pm\uq$ for $\delta=0.4$ and $a_1=1$ (so $\gamma_1 \approx 0.324$ in \eqref{gkak}). The lower bound of \eqref{cl1} is approximately $0.6$ and 
	the upper bound is about $9.8$, while the upper bound in \eqref{dec2} is approximately $38.9$. For a fixed $\delta$, a different pressure behavior is obtained by varying omega.  \textit{Left}: for $\omega^2=1<38.9$ 
	the minimum (in blue) is at $(-\uq,0)$ and the maximum (in red) is at $(\uq,0)$. \textit{Right:} for $\omega^2 \approx 39.1$ the minimum is at $(\uq,y_0)$ and the maximum is at $(-\uq,y_0)$ with $y_0 \approx -0.42$. }
	\label{fig:pdyn}
\end{figure}

\begin{figure}[htbp]
  \centering
  \includegraphics[width=0.85\textwidth]{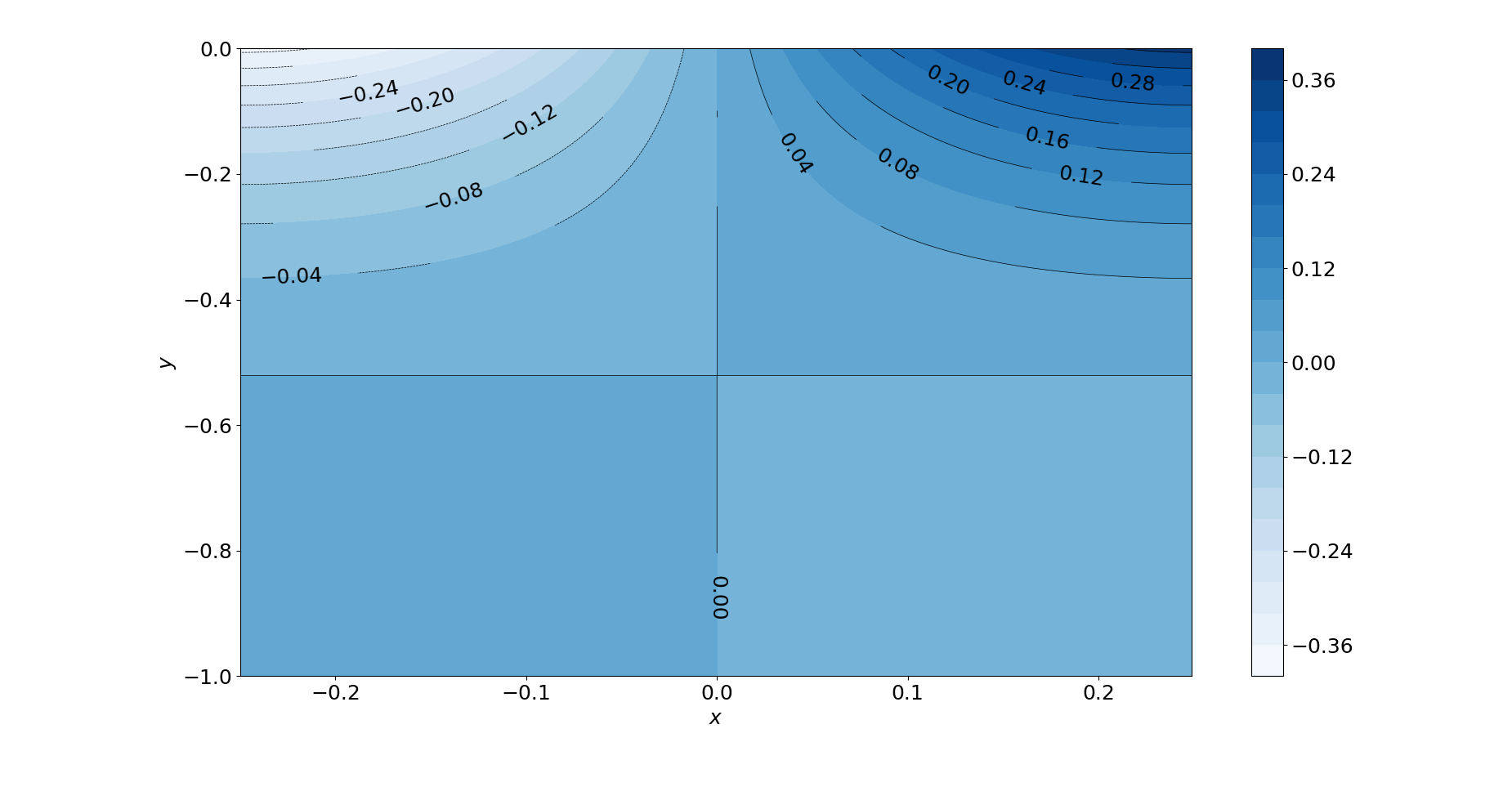}
  \includegraphics[width=0.85\textwidth]{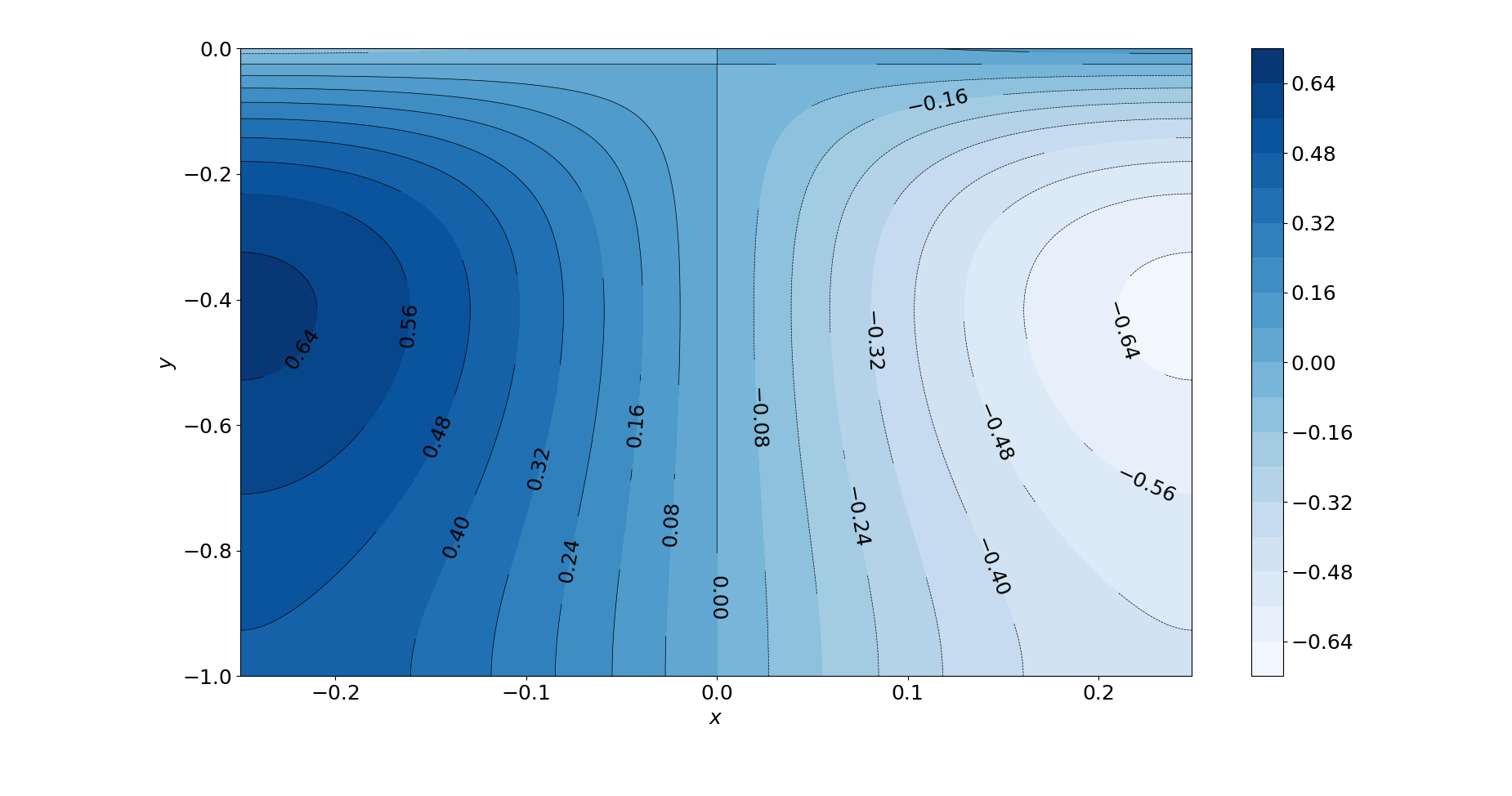}
    \caption{Contour map of the dynamic pressure from Fig. 7: above for $\omega=1$ and below for $\omega=\sqrt{39.1}$.}
    \label{fig:cm}
\end{figure}

Figure \ref{fig:pdyn} illustrates one case when \eqref{cl2} holds, and consequently the minimum of the dynamic pressure is at $(\uq,y_0)$ and the maximum at $(-\uq,y_0)$, 
and one case when \eqref{cl1} and \eqref{dec2} hold simultaneously, so that the minimum is at $(-\uq,0)$ and the maximum at $(\uq,0)$. The corresponding  contour maps 
are depicted in figure \ref{fig:cm}.

\section{Hydrodynamic pressure}

In Section \ref{sec:dyn} we have investigated the distribution of the dynamic pressure. The goal of the present section is to present a detailed study of the hydrodynamic pressure.

Within the framework of linear theory, the general expression for the total pressure beneath a surface wave of type \eqref{lw} propagating to the right with amplitude \eqref{A} is 
the sum of the hydrostatic pressure and the dynamic pressure, given by
\begin{align}
P =& P_{atm}-\rho g d y \label{hydp} \\
&- \epsilon g \rho d \delta \gamma_1 \Big([\hc-\omega (1+y)]\cosh[2\pi \delta (y+1)]+\f{\omega}{2 \pi \delta} \sinh[2 \pi \delta(y+1)]\Big)\sin(2\pi x)\,.  \nonumber
\end{align}
Given that $\varepsilon \ll 1$, we have $P_y <0$ for $y \in [-1,0]$, so that the hydrodynamic  pressure increases with depth. Note that the underlying constant vorticity influences the behaviour of the pressure at 
every fixed depth $y$, which is determined by the features of the dynamic pressure that were discussed in detail in the previous section. At every fixed depth, the extrema of the hydrodynamic pressure 
occur on the crest line $x=\frac{1}{4}$ and on the trough line $x=-\frac{1}{4}$, and the analysis performed in the Section 3 enable us conclude that the only major qualitative difference is between the cases without 
flow-reversal and with flow-reversal:
\begin{itemize}
\item In the case without flow reversal is the hydrodynamic pressure strictly monotone between the crest line and the trough line, with the maximum on the crest line $x=\frac{1}{4}$ and the minimum on 
the trough line $x=-\frac{1}{4}$.
\item As shown in Section 3, flow reversal for right-propagating waves is only possible for a positive vorticity $\omega>0$ subject to the constraint specified in \eqref{frl}. From  \eqref{hydp} we see that along the flat bed $y=-1$ we have
$$P=P_{atm} + \rho gd - \epsilon g \rho d \delta \gamma_1\,\hc \,\sin(2\pi x)$$
with $\hc>0$ and $\gamma_1>0$ (see the discussion at the beginning of Section 3.5). Consequently, along the flat bed the hydrodynamic pressure is strictly monotone between the crest line and the trough line, with the minimum 
below the crest, at $x=\frac{1}{4}$, and the maximum below the trough, at $x=-\frac{1}{4}$. The discussion in Section 3.5 shows that this behaviour changes at the depth level $y=y_+ \in (-1,0)$ determined by the unique solution 
$y_+ \in (y_0,0)$ of equation \eqref{y+}, where $y=y_0 \in (-1,0)$ is the critical level. Along every level $y>y_+$ the maximum of the hydrodynamic pressure is attained below the wave crest and the minimum below the 
wave trough, while along a level $y < y_+$ the locations of the extrema flip, the minimum of the hydrodynamic pressure being attained below the wave crest and the maximum occurring below the 
wave trough (see Fig. 6 for a depiction of the overall situation).
\end{itemize}

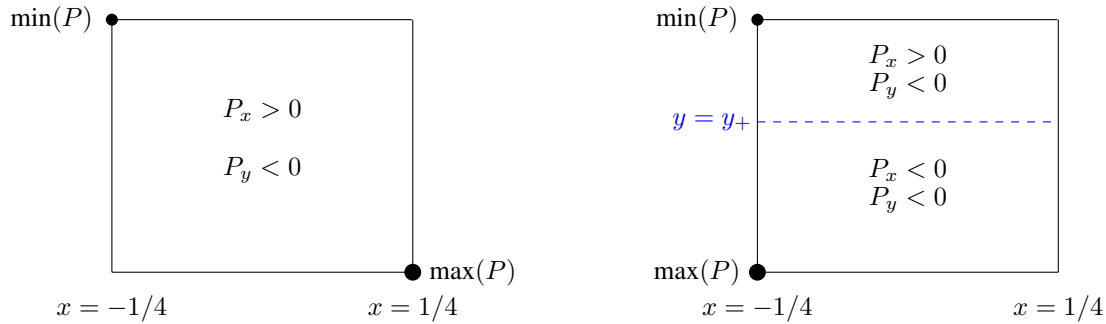
\begin{figure}[htbp]
	\centering
	\begin{minipage}{0.49 \textwidth}
	\centering
		\begin{tikzpicture}[scale=8]
\draw[smooth,samples=100,domain=1.25:1.75] plot(\x,0);
\draw (1.75,-0.42) -- (1.75,0);
\draw (1.25,0)-- (1.25,-0.42);
\draw (1.25,-0.42)-- (1.75,-0.42);
			\draw(1.15,0) node{min$(P)$};
			\draw(1.88,-0.42) node{max$(P)$};
				\fill(1.75,-0.42) circle(0.02);
				\fill(1.25,0) circle(0.01);
			\draw(1.5,-0.15) node{$P_x>0$};
			\draw(1.5,-0.25) node{$P_y<0$};
			\draw(1.25,-0.48) node{$x=-1/4$};
			\draw(1.75,-0.48) node{$x=1/4$};
		\end{tikzpicture}
	\end{minipage}
	\begin{minipage}{0.49 \textwidth}
	\centering
		\begin{tikzpicture}[scale=8]
\draw[smooth,samples=100,domain=1.25:1.75] plot(\x,0);
\draw (1.75,-0.42) -- (1.75,0);
\draw (1.25,0)-- (1.25,-0.42);
\draw (1.25,-0.42)-- (1.75,-0.42);
			\draw(1.15,0) node{min$(P)$};
			\draw(1.12,-0.42) node{max$(P)$};
				\fill(1.25,-0.42) circle(0.02);
				\fill(1.25,0) circle(0.01);
			\draw(1.5,-0.06) node{$P_x>0$};
			\draw(1.5,-0.11) node{$P_y<0$};
			\draw(1.5,-0.25) node{$P_x<0$};
			\draw(1.5,-0.30) node{$P_y<0$};
			\draw(1.25,-0.48) node{$x=-1/4$};
			\draw(1.75,-0.48) node{$x=1/4$};
				\draw[blue,dashed] (1.25,-0.17)--(1.75,-0.17) node[pos=-0.15]{$y=y_+$};
		\end{tikzpicture}
	\end{minipage}
	\caption{Minimum and maximum of the hydrodynamic pressure. \textit{Left}: The case without flow-reversal. \textit{Right:} The flow-reversal flow case, with the value $y^+$ defined by $P_x(x,y^+)=0$ with $y^+$ the unique solution of equation \eqref{y+} in $[-1,0)$.}
	\label{fig:hydrodyn}
\end{figure}

The figure \ref{fig:hydrodyn} illustrates the two possible cases (with or without flow-reversal) of the locations of the maximum and minimum of the hydrodynamic pressure.

\section{Conclusion}

We have presented a detailed study of the dynamic and hydrodynamic pressure beneath a periodic travelling-wave propagating in constant-vorticity flow at the surface of water with a flat bed. While the 
hydrostatic pressure dominates the dynamic pressure, the latter, being attuned to the impact of fluid movement, is more sensitive to the effects of vorticity and may therefore be used to detect the 
presence of non-uniform underlying currents. The most significant vorticity effects are noticeable in the case of flow-reversal, when the locations of the extrema of the dynamic pressure can be very different from those 
typical for irrotational flow (in which case the maximum is attained at the wave crest and the minimum at the wave trough): depending on the strength of the vorticity, the extrema may occur on the flat bed or on the critical level. 
The vorticity effect on the hydrodynamic pressure manifests itself in altering the location of the extrema at a fixed depth level in the case of flow-reversal. In particular, the maxima/minima locations along the flat bed are flipped with 
respect to those along the free surface.

\subsection*{Acknowledgements}
This research was funded in part by the Austrian Science Fund
(FWF) Z 387-N (AC) and 10.55776/P34981 (OS) -- New Inverse Problems of Super-Resolved Microscopy (NIPSUM),
SFB 10.55776/F68 (OS) ``Tomography Across the Scales'', project F6807-N36
(Tomography with Uncertainties), and 10.55776/T1160 (OS) ``Photoacoustic Tomography: Analysis and Numerics''. For open access purposes, the author has applied a CC BY public copyright license to any author-accepted manuscript version arising from this submission. The financial support by the Austrian Federal Ministry for Digital and Economic
Affairs, the National Foundation for Research, Technology and Development and the Christian Doppler
Research Association is gratefully acknowledged. The authors
are grateful for helpful comments from the referees.

\subsection*{Data Availability} All data for this paper are properly cited and referred to: the relevant data can be found in
 \cite{cb, e, eli, k, m, z}.

\subsection*{Conflict of interest} The content of this paper is original and has not been published or submitted for publication
elsewhere. The authors also declare that they have no Conflict of interest.

\section*{Appendix 1: Decomposition in Fourier modes}

The time-dependence for travelling waves amounts to a simple translation 
of the horizontal spatial variable $x$ by $\hat{c}\,t$, so that it suffices to investigate the problem \eqref{lint} at time $t=0$.
Note that the first and last relation in \eqref{lint} yield
\begin{linenomath*}
\begin{equation}\label{ndv}
\delta^2\,\hat{v}_{xx} + \hat{v}_{yy}=0\quad\hbox{for}\quad -1 < y < 0\,.
\end{equation}
\end{linenomath*}
Seeking smooth solutions, the Fourier series expansion
\begin{linenomath*}
$$\hat{v}(x,y)=\sum_{k \in {\mathbb Z}} \alpha_k(y)\,e^{2\pi i kx},\quad -1 \le y \le 0\,,$$
\end{linenomath*}
in (\ref{ndv}) leads to 
\begin{linenomath*}
$$\alpha_k''(y)-4\pi^2k^2\delta^2\,\alpha_k(y)=0\,,\qquad y \in (-1,0)\,,$$
\end{linenomath*}
for every $k \in {\mathbb Z}$. For $k \neq 0$ we obtain that 
\begin{linenomath*}
$$\alpha_k(y)=a_k\,e^{2\pi k \delta y}+b_k\,e^{-2\pi k \delta y},\qquad -1 \le y \le 0\,,$$
\end{linenomath*}
while for $k=0$ we have $\alpha_0(y)=a_0 y +b_0$; here $a_k,\,b_k \in {\mathbb R}$ are some 
constants.  The boundary condition on $y=-1$ in (\ref{lint}) forces $a_0=b_0$ and 
$b_k=-a_k\,e^{-4\pi k \delta}$ for every $k \in {\mathbb Z} \setminus \{0\}$, so that
\begin{linenomath*}
$$\hat{v}(x,y)=a_0(y+1)+\sum_{k \in {\mathbb Z} \setminus \{0\}} 2\,a_k\,e^{-2\pi k \delta}\sinh[2\pi k \delta(y+1)]\,e^{2\pi i kx},\quad -1 \le y \le 0\,.$$
\end{linenomath*}
Since $\hat{v}$ has to be a real-valued function, for $k \neq 0$ we must have 
$$a_k e^{-2\pi k \delta}\sinh[2\pi k \delta(y+1)]\,\sin{2\pi kx} + a_{-k} e^{2\pi k \delta}\sinh[2\pi k \delta(y+1)]\,\sin{2\pi  kx}=0\,,$$
that is,
\begin{linenomath*}
$$a_{-k}\,e^{2\pi k \delta}=-\,a_k\,e^{-2\pi k \delta}\quad\text{for all}\quad k \in {\mathbb Z} \setminus \{0\}\,.$$
\end{linenomath*}
Setting 
\begin{linenomath*}
\begin{equation}\label{gkak}
\gamma_k=4a_k\,e^{-2\pi k \delta}\,,\qquad k \in {\mathbb Z} \setminus \{0\}\,,
\end{equation} 
\end{linenomath*}
we get
\begin{linenomath*}
$$\hat{v}(x,y)=a_0(y+1)+\sum_{k \ge 1} \gamma_k\,\sinh[2\pi k \delta(y+1)]\,\cos(2\pi kx)\,.$$
\end{linenomath*}
The first equation in \eqref{lint} yields
$$\frac{{\rm d}}{{\rm d}y} \int_0^1 \hat{v}(x,y)\,{\rm d}x = - \int_0^1 \frac{{\rm d}\,\hat{u}}{{\rm d}x}(x,y)\,{\rm d}x=0$$
due to the periodicity in the $x$-variable. Taking into account the fourth relation in \eqref{lint}, we see that 
the mean of $\hat{v}$ vanishes at every level $y \in [-1,0]$. Consequently, we must have $a_0=0$, so that
\begin{linenomath*}
\begin{equation}\label{vnd}
\hat{v}(x,y)=\sum_{k \ge 1} \gamma_k\,\sinh[2\pi k \delta(y+1)]\,\cos(2\pi kx)\,.
\end{equation} 
\end{linenomath*}
Combining the first and last relation in (\ref{lint}), we now infer that
\begin{linenomath*}
$$\hat{u}(x,y)=\gamma_0- \delta\,\sum_{k \ge 1} \gamma_k\,\cosh[2\pi k \delta(y+1)]\,\sin(2\pi kx)\,,$$
\end{linenomath*}
for some constant $\gamma_0 \in {\mathbb R}$ which represents the mean flow over the flat bed $y=-1$. Assuming zero mean on the flat bed $y=-1$ means $\gamma_0=0$, so that
\begin{linenomath*}
\begin{equation}\label{und}
\hat{u}(x,y)=- \delta\,\sum_{k \ge 1} \gamma_k\,\cosh[2\pi k \delta(y+1)]\,\sin(2\pi kx)\,.
\end{equation}
\end{linenomath*}
Note that 
\begin{linenomath*}
\begin{equation}\label{psiuv}
\hat{u}=\hat{\psi}_y\,,\qquad \hat{v}=-\hat{\psi}_x\,, \qquad x \in {\mathbb R}\,, \quad -1 \le y \le 0\,,
\end{equation}
\end{linenomath*}
where 
\begin{linenomath*}
\begin{equation}\label{sli}
\hat{\psi}(x,y)=- \sum_{k \ge 1} \frac{\gamma_k}{2\pi k}\,\sinh[2\pi k \delta(y+1)]\,\sin(2\pi kx)\,,
\end{equation}
\end{linenomath*}
is the stream function. Therefore the second equation in (\ref{lint}) yields
\begin{linenomath*}
$$\hat{p}(x,y)=[\hat{c}-\omega(1+y)]\,\hat{u}(x,y) + \omega \hat{\psi}(x,y) +\beta(y)\,,\qquad y \in [-1,0]\,,$$
\end{linenomath*}
for some function $\beta$. Using now the last equation of (\ref{lint}) in the third relation 
of (\ref{lint}), we see that $\beta$ is a constant. Thus
\begin{linenomath*}
\begin{equation}\label{dpl}
\hat{p}(x,y)=[\hat{c} - \omega(1+y)]\,\hat{u}(x,y) + \omega \hat{\psi}(x,y) +\beta\,.
\end{equation}
\end{linenomath*}

From the sixth relation in (\ref{lint}) and (\ref{und}) we now obtain
\begin{linenomath*}
$$h(x) = \beta -\,\delta [\hat{c}-\omega]\sum_{k \ge 1} \gamma_k\, \cosh(2\pi k\delta)\,\sin(2\pi kx)\\
- \omega \sum_{k \ge 1} \frac{\gamma_k}{2\pi k}\, \sinh(2\pi k\delta)\,\sin(2\pi kx) \,.$$
\end{linenomath*}
Since the average of $h$ over one period should vanish, as it represents the mean water level $y=0$, we must have 
\begin{linenomath*}
\begin{equation}\label{constlin}
\beta=0\,,
\end{equation} 
\end{linenomath*}
so that
\begin{linenomath*}
\begin{equation}\label{h1}
h(x)=- \,\delta [\hat{c}-\omega]\sum_{k \ge 1} \gamma_k\, \cosh(2\pi k\delta)\,\sin(2\pi kx)- 
\omega \sum_{k \ge 1} \frac{\gamma_k}{2\pi k}\, \sinh(2\pi k\delta)\,\sin(2\pi kx)\,.
\end{equation}
\end{linenomath*}
Differentiation yields, in view of the fifth relation in (\ref{lint}) and (\ref{vnd}), 
\begin{linenomath*}
$$ \sum_{k \ge 1} \,\gamma_k\,\Big\{ 2\pi k\delta [\hat{c}-\omega]^2 \cosh(2\pi k\delta) + 
\Big(\omega [\hat{c}-\omega] - 1\Big)\sinh(2\pi k\delta)\Big\}\, \cos(2\pi kx) =0\,.$$
\end{linenomath*}
Consequently
\begin{linenomath*}
\begin{equation}\label{eqfc1}
\gamma_k\,\Big\{  [\hat{c}-\omega]^2  + 
\frac{\omega\tanh(2\pi k\delta)}{2\pi k \delta}\, [\hat{c}-\omega] - \frac{\tanh(2\pi k\delta)}{2\pi k \delta}\Big\}=0\,,\qquad k \ge 1\,.
\end{equation}
\end{linenomath*}
If $\gamma_{k} \neq 0$ for some $k \ge 1$, then \eqref{eqfc1} becomes a quadratic polynomial equation in $[\hat{c}-\omega]$ and we obtain the \textit{dispersion relation} 
\begin{linenomath*}
\begin{equation}\label{eqfc2}
\hat{c}-\omega =-\,\frac{\omega\tanh(2\pi k\delta)}{4\pi k \delta} \pm \sqrt{\frac{\omega^2\tanh^2(2\pi k \delta)}{(4\pi k \delta)^2} + 
\frac{\tanh(2\pi k \delta)}{2\pi k \delta}}\,.
\end{equation}
\end{linenomath*}
For $s>0$, set
\begin{linenomath*}
$$S_\pm(s)=-\frac{\omega s}{2} \pm \sqrt{\frac{\omega^2 s^2}{4} + s}\,,$$
\end{linenomath*}
chosen so that $S_-(s) <0<S_+(s)$ for $s=\frac{\tanh(2\pi k \delta)}{2\pi k \delta}$.  Since
$$\left( 1+\frac{\omega^2 s}{2}\right)^2 = 1+\f{s^2 \omega^4}{4} +s \omega^2   > \f{s^2 \omega^4}{4}+s \omega^2=\left( \omega\sqrt{\frac{\omega^2 s^2}{4} + s} \right)^2\,,$$
we see that the function $s \mapsto S_+(s)$ is strictly increasing for $s>0$, while $s \mapsto S_-(s)$ is strictly decreasing. 
Consequently, due to \eqref{eqfc2}, there is at most one integer $k \ge 1$ for which $\gamma_k \neq 0$, meaning that \textit{superpositions of Fourier 
modes are not possible}. Indeed, due to monotonicity, there are no integers $k_2>k_1 \ge 1$ 
such that $S_+(2\pi k_2\delta)=S_+(2\pi k_1\delta)>0$ or $S_-(2\pi k_2\delta)=S_-(2\pi k_1\delta)<0$.

\section*{Appendix 2: Some properties of the dynamic pressure for flows with nonzero vorticity}

In Section 2.1.3, to compare the four values of the dynamic pressure given in \eqref{lptop}-\eqref{lpbed}, we claimed that elementary calculations show that \eqref{clp3} holds unless $\omega^2$ 
exceeds the lower bound given in \eqref{bol}. Since these calculations are somewhat intricate, we provide the details. 

To understand how the four values in \eqref{lptop}-\eqref{lpbed} relate to each other note that, using the dispersion relation \eqref{eqfc2}, we can write
\begin{linenomath*}
$$\frac{\hat{c}}{[\hat{c}-\omega]\,\cosh[2\pi \delta] + 
\frac{\omega}{2\pi \delta}\,\sinh[2\pi \delta]}=\frac{\omega 
-\,\frac{\omega\tanh(2\pi \delta)}{4\pi  \delta} \pm \sqrt{\frac{\omega^2\tanh^2(2\pi  \delta)}{(4\pi  \delta)^2} + 
\frac{\tanh(2\pi  \delta)}{2\pi  \delta}}}{\cosh[2\pi \delta]\Big\{\frac{\omega\tanh(2\pi \delta)}{4\pi  \delta} 
\pm \sqrt{\frac{\omega^2\tanh^2(2\pi  \delta)}{(4\pi  \delta)^2} + 
\frac{\tanh(2\pi  \delta)}{2\pi  \delta}}\Big\}}\,,$$
\end{linenomath*}
which, for $\omega \neq 0$, takes the form
\begin{linenomath*}
\begin{equation}\label{clp2}
\frac{\hat{c}}{[\hat{c}-\omega]\,\cosh[2\pi \delta] + 
\frac{\omega}{2\pi \delta}\,\sinh[2\pi \delta]} = \frac{1 
-\,\frac{\tanh(2\pi \delta)}{4\pi  \delta} \pm \frac{\omega}{|\omega|} \sqrt{\frac{\tanh^2(2\pi  \delta)}{(4\pi  \delta)^2} + 
\frac{\tanh(2\pi  \delta)}{2\pi  \delta \omega^2}}}{\cosh[2\pi \delta]\Big\{\frac{\tanh(2\pi \delta)}{4\pi  \delta} 
\pm  \frac{\omega}{|\omega|} \sqrt{\frac{\tanh^2(2\pi  \delta)}{(4\pi  \delta)^2} + 
\frac{\tanh(2\pi  \delta)}{2\pi  \delta \omega^2}}\Big\}}\,,
\end{equation}
\end{linenomath*}

Let us now verify the claim that \eqref{clp3} holds unless $\omega$ satisfies \eqref{bol}. 

Indeed, for $\omega=0$ the expression in the middle of \eqref{clp3} equals to $1/\cosh[2\pi \delta] \in (0,1)$. On the other hand, for 
$\omega \neq 0$, due to \eqref{clp2}, \eqref{clp3}  is equivalent to the inequality
\begin{linenomath*}
\begin{equation}\label{clp4}
-1< \frac{1- \frac{\tanh(s)}{2s} \pm  \frac{\omega}{|\omega|} \ \sqrt{\frac{\tanh^2(s)}{4s^2}+\frac{\tanh(s)}{s\omega^2}}}{ \cosh(s)\Big\{\frac{\tanh(s)}{2s} 
\pm  \frac{\omega}{|\omega|} \sqrt{\frac{\tanh^2(s)}{4s^2}+\frac{\tanh(s)}{s\omega^2}}\Big\}  }<1 \quad\text{for}\quad s=2\pi\delta>0\,.
\end{equation}
\end{linenomath*}
For the choice of the $+$ sign in \eqref{clp4} we have:
\begin{itemize}
\item If $\omega>0$, then the expression to be estimated in \eqref{clp4} is positive, so that the lower bound in \eqref{clp4} is trivial, while the validity of the upper bound is equivalent to 
\begin{linenomath*}
$$1 < \frac{\tanh(s)}{2s} [1+\cosh(s)] + [\cosh(s)-1] \sqrt{ \frac{\tanh^2(s)}{4s^2}+ \frac{\tanh(s)}{s\omega^2}}\,.$$
\end{linenomath*}
This holds true since the right side above exceeds
\begin{linenomath*}
$$\frac{\tanh(s)}{2s} [1+\cosh(s)] + [\cosh(s)-1] \frac{\tanh(s)}{2s}=\frac{\sinh(s)}{s} >1\,.$$
\end{linenomath*}
\item If $\omega < 0$, since the denominator of the expression to be evaluated in \eqref{clp4} is negative, the validity of the upper/lower bound is equivalent to
\begin{linenomath*}
$$1+  [\cosh(s)-1] \sqrt{ \frac{\tanh^2(s)}{4s^2}+ \frac{\tanh(s)}{s\omega^2}} > [\cosh(s)+1] \,\frac{\tanh(s)}{2s}$$
\end{linenomath*}
and
\begin{linenomath*}
$$[\cosh(s)+1] \sqrt{ \frac{\tanh^2(s)}{4s^2}+ \frac{\tanh(s)}{s\omega^2}} > 1+  [\cosh(s)-1]\,\frac{\tanh(s)}{2s} \,,$$
\end{linenomath*}
respectively. The first inequality above holds because its left side exceeds
\begin{linenomath*} 
$$1+  [\cosh(s)-1] \,\frac{\tanh(s)}{2s} > [\cosh(s)+1] \,\frac{\tanh(s)}{2s}$$
\end{linenomath*}
since $1 >\frac{\tanh(s)}{s}$. As for the second inequality above, by squaring both its positive sides, we see that it is equivalent to 
\begin{linenomath*} 
\begin{equation}\label{sshear}
\omega^2 < \frac{ [\cosh(s)+1]^2 \frac{\tanh(s)}{s}}{ 1 + [\cosh(s)-1] \frac{\tanh(s)}{s} - \cosh(s)\,\frac{\tanh^2(s)}{s^2}}\,,
\end{equation}
\end{linenomath*}
which is precisely the opposite of \eqref{bol}. 
\end{itemize}
For the choice of the $-$ sign in \eqref{clp4} we have again the two previously discussed cases, since the $\pm$ sign for $\omega$ corresponds precisely to the 
$\mp$ sign for $-\omega$. We thus verified \eqref{clp4} if \eqref{bol} holds.

The above discussion shows also how the expression in the middle of \eqref{clp4} behaves if \eqref{bol} fails: for the choice of the $+$ sign in \eqref{clp4}, the inequality remains valid for 
$\omega>0$ but the lower bound becomes invalid if $\omega <0$. The fact that one can switch between the $\pm$ signs in \eqref{clp4} by changing $\omega$ to $-\omega$ means that 
for the choice of the $-$ sign in \eqref{clp4}, the inequality remains valid for 
$\omega<0$ but the lower bound becomes invalid if $\omega >0$.

\end{document}